\documentclass[12pt, aps, prd, superscriptaddress, nofootinbib, 
               amsmath, amssymb,
               preprintnumbers, showpacs,
               raggedbottom,
               floatfix]{revtex4-1}

\usepackage[T1]{fontenc}
\usepackage{textcomp}
\usepackage{etoolbox}
\usepackage{etextools}
\usepackage{comment}

\usepackage[acronym]{glossaries}
\glsdisablehyper
\makeglossaries
\usepackage[utf8]{inputenc}

\usepackage{hyperref}
\usepackage{graphicx}
\usepackage{dcolumn}
\usepackage{color}

\def  \bea  {\begin{eqnarray}}
\def  \eea  {\end{eqnarray}}
\def  \nn   {\nonumber}
\def \l {\left}
\def \r {\right}

\newcommand{\stkout}[1]{\ifmmode\text{\sout{\ensuremath{#1}}}\else\sout{#1}\fi}

\begin{document}
\title{MURCA driven Bulk viscosity in neutrino trapped baryonic matter}
\author{Sreemoyee Sarkar}
\email{sreemoyee.sarkar@nmims.edu}
\affiliation{Mukesh Patel School of Technology Management and Engineering, SVKM’s NMIMS University, Vile Parle (W), Mumbai 400056, India}

\author{Rana Nandi}
\email{rana.nandi@snu.edu.in}
\affiliation{Department of Physics, School of Natural
Sciences, Shiv Nadar Institution of Eminence, Greater Noida
201314, Uttar Pradesh, India }

\date{\today}

\begin{abstract}
We examine bulk viscosity, taking into account trapped neutrinos in baryonic matter, in the context of binary neutron star mergers. Following the merging event, the binary star can yield a  remnant compact object with densities up to $5$ nuclear  saturation density and temperature upto $50$ MeV resulting in the retention of neutrinos. We employ two relativistic mean field models, NL3 and DDME2, to describe the neutrino-trapped baryonic matter. The dissipation coefficient is determined by evaluating the Modified URCA interaction rate in the dense baryonic medium, and accounting for perturbations caused by density oscillations.  We observe the resonant behavior of bulk viscosity as it varies with the temperature of the medium. The bulk viscosity peak remains within the temperature range of $\sim 13-50$ MeV, depending upon the underlying equation of states and lepton fractions. This temperature range corresponds to the relevant domain of binary neutron star mergers. We also note that in presence of neutrinos in the medium the bulk viscosity  peak shifts  towards higher temperature and  the  peak value of bulk viscosity also changes. The time scale of viscous dissipation is dictated by the beta-off-equilibrium susceptibilities derived from the nuclear equation of state. The resulting viscous decay time scale ranges from   $32-100$ milliseconds, which aligns with the order of magnitude of the post-merger object's survival time in some specific scenarios. 
\end{abstract}


\maketitle
\section{Introduction}
The detection of gravitational waves by the LIGO-VIRGO detector \cite{LIGOScientific:2016aoc, LIGOScientific:2017vwq} has brought significant interest in studying matter under extreme conditions. In the event of a binary neutron star (BNS) merger, a compact object with density several times nuclear saturation density ($n_0 = 0.16$ fm$^{-3}$), and temperatures ($T$) up to several tens of  MeV can be formed. Post merging, if the mass of the compact object is larger than the Tolman–Oppenheimer–Volkoff (TOV) mass, it collapses to form a black hole within hundred milliseconds \cite{Shibata:1999wm, Baiotti:2008ra}. If the remnant object is less massive than the TOV limit, it survives as a neutron star. Efforts to numerically simulate the merger scenario by employing Einstein’s theory of general relativity (GRHD) were started  decades before detecting the first binary neutron star merger event, GW170817 \cite{LIGOScientific:2017vwq, Faber:2012rw, East:2015vix, PhysRevLett:107:051102, PhysRevD:93:044019, PhysRevD:96:043019}. 
%

Immediately after merging, the nuclear fluid in the merger remnant experiences wild density oscillations. These oscillations get damped by a dissipative process if the relevant timescale is in the same order  of hundred milliseconds. This time scale is determined by the thermodynamics of the background medium and the kinematics of the relevant processes. To assess the importance of a particular transport process in the simulation, one needs to determine the timescale on which it acts. If the timescale of the particular transport process is comparable to the merger timescale, i.e., hundred milliseconds, then that dissipative process is considered relevant for the simulation of the merger. In Ref.\cite{Alford:2017rxf} it has been shown that  the  Bulk viscosity ($\zeta$) of the hadronic medium plays an important role in controlling density oscillation. Along with  this recently in Refs.\cite{Harutyunyan:2018mpe, Sarkar:2023fwu} the importance of other dissipative processes  like electron-transport coefficients in
a magnetized high temperature and high density electron-ion plasma, in the
context of simulations for BNS merger, has also been explored.  In all the recent studies on  dissipative processes  in the context of BNS mergers \cite{Most:2022yhe, Sedrakian:2022kgj, Alford:2021lpp,  Alford:2019qtm,Celora:2022nbp, Alford:2022ufz, Alford:2018lhf, Alford:2019kdw, Most:2021zvc, Harris:2020rus, Alford:2021ogv,   Alford:2023gxq, Alford:2023uih, Camelio:2022fds, Camelio:2022ljs}, the calculations have been extended to the temperature domain, reaching several tens of MeV. At this extremely high temperature regime, the mean free path of neutrinos is smaller than the size of the stellar object. Thus, neutrinos remain trapped inside the matter, resulting in a non-zero neutrino chemical potential.  Once temperature starts to decrease neutrinos start to escape from the medium to make the baryonic matter free of neutrinos.

Vibrational and rotational instabilities in the merged object disrupts the state of beta equilibrium within the baryonic matter. The rate at which this deviation in beta equilibrium converges to zero provides insight into how quickly the  particle concentrations adapt to pressure variations resulting from compression and rarefaction. The alteration in particle concentrations is initiated by pressure fluctuations, resulting in the development of a non-zero difference in chemical potentials between the initial and final states of interacting particles, denoted by  $\mu_{\Delta}$. Non-zero $\mu_{\Delta}$ serves as an indicator of the deviation from beta equilibrium. The interacting particles within the baryonic medium comprise neutrons, protons, as well as leptons such as electrons and neutrinos, involving processes like direct URCA(DURCA) and modified URCA (MURCA). Through these electroweak processes the system responds to the  perturbation caused due to density oscillation. It's important to note that these electroweak processes have significant impact on bulk viscosity since their time scales align closely with the oscillation frequency of the compact object.

A series of recent works (\cite{Most:2022yhe, Sedrakian:2022kgj, Alford:2021lpp,  Alford:2019qtm, Celora:2022nbp, Alford:2022ufz, Alford:2018lhf, Alford:2019kdw, Most:2021zvc, Harris:2020rus, Alford:2021ogv,  Alford:2023gxq, Alford:2023uih}) addressed various microphysical aspects of bulk viscosity in BNS mergers. 
These studies explored several key aspects, including the impact of weak interaction in bulk viscosity (\cite{Most:2021zvc}), the rate of beta equilibration (\cite{Alford:2021ogv}), the influence of trapped neutrinos on interaction rates (\cite{Alford:2019kdw, Alford:2023gxq, Alford:2021lpp}), the effects of hyperons in neutron star mergers (\cite{Sedrakian:2022kgj}), and the considerations of isospin equilibration in neutron star mergers (\cite{Alford:2023gxq}). A comprehensive analysis of the damping of density oscillations in the neutrino-transparent matter has also been performed (\cite{Alford:2019qtm, Alford:2023uih}. Furthermore, some studies discussed various methods of implementing bulk viscosity in a BNS merger simulation (\cite{Celora:2022nbp, Camelio:2022fds, Camelio:2022ljs}). Very recently, the first binary neutron star simulation that considered the dissipative effects self-consistently has shown that large bulk viscosity can significantly damp the oscillations of the stellar cores just after the merger and, as a result, can substantially affect the characteristics of the post-merger gravitational wave signals \cite{Chabanov:2023blf}.

Before the detection of Gravitational waves, the calculation of bulk viscosity was focused on isolated neutron stars containing either nuclear matter or quark matter \cite{Madsen:1992sx, PhysRevD.39.3804, Haensel:2001mw, Jones:2001ya, Alford:2006gy, Dong:2007ax,PhysRevD.45.4708, PhysRevD.76.083001}). All such calculations were performed within a regime where the amplitude of
variations of the off-equilibrium chemical potential
 is significantly smaller than the system’s temperature. Within this regime, the system exhibits a linear response to changes in pressure. However, authors in Ref.\cite{Alford:2010gw} first attempted to investigate the significance of large amplitude oscillations in bulk viscous dissipation for isolated neutron stars.

       In light of recent research on binary neutron star mergers, as mentioned earlier, we formulate $\zeta$ pertaining to the MURCA process with trapped neutrinos. This formulation is designed to accommodate scenarios where the MURCA interaction rate is a nonlinear function of the perturbation, specifically, a nonlinear function of $\mu_{\Delta}$.  To compute $\zeta$, we solve  integro-differential equation of the chemical potential fluctuation to obtain a general solution of $\mu_{\Delta}$. We then apply a weighting factor of $\cos \omega t$ ($\omega$ represents the oscillation frequency of the merged object and $t$ denotes time) and integrate over a single time period during the compression and rarefaction cycle. The current calculation holds significance on two fronts. Firstly, within this paper, we incorporate the MURCA interaction rate in the presence of trapped neutrinos to evaluate bulk viscous dissipation in BNS mergers. Secondly, our formulation provides a general approach to obtaining the perturbation, characterized by $\mu_{\Delta}$, which leads to bulk viscous dissipation.

        The paper is organized as follows in Section: II we discuss the formalism of bulk viscosity of hadronic matter in presence of trapped neutrinos.  In Section III, we present numerical estimation of $\zeta$ and variation of it with different parameters. Finally, in Section IV, we  summarize and conclude.
\section{Bulk Viscosity of trapped neutrino dense matter}
Bulk viscosity arises as a response to a system undergoing a repetitive cycle of compression and rarefaction. This cyclic behavior of compression and rarefaction induces density oscillations in conserved quantities such as the baryon number density, denoted as $n_{ B}(\vec r,t) = \bar{n}_{B}(\vec r)+\delta n_{B}(\vec r,t)=\bar{n}_{B}(\vec r)+\Delta n_{B}(\vec r,t)\rm sin(\omega t)$. Here, $\bar{n}_{B}$ represents the equilibrium value,  $\delta n_{B}$ represents the harmonic oscillation component, $\Delta n_B$ is the amplitude of the oscillation and $\omega$ is the frequency of density oscillation. The oscillatory behavior  impacts the rate of beta equilibration, resulting in an asymmetry between the rates of forward and backward reactions for weak interaction processes. By subtracting the initial state's chemical potential from the final state's chemical potential, we identify the difference $\mu_{\Delta}$ as a specific quantity that acts as a perturbation to the bulk viscosity.  In our calculation, we consider weak interaction processes due to their time scale becoming comparable to the  rotation period of star. On the other hand, the contribution of strong interaction to bulk viscosity calculations is considered negligible, as the re-equilibration time scale does not align with the oscillation period of the star.

For a particular weak process, we get $\mu_{\Delta}$ by subtracting the final state chemical potential from the initial state :   
\bea
\mu_{\Delta}=\sum_i\mu_i-\sum_f \mu_f.
\eea
 $\mu_{\Delta}$ 
is non-zero due to density fluctuations, and re-equilibration of this  quantity leads to bulk viscosity. Since the equilibrium state can be described by the baryon density and proton fraction ($x_p$), the fluctuations in $\mu_\Delta$ can be written as,
 \bea
 \delta\mu_\Delta=
\left.\frac{\partial \mu_\Delta}{\partial n_{B}} \right|_{x_p} \delta n_{B}
+\left.\frac{\partial \mu_\Delta}{\partial x_p}\right|_{n_{B}} \delta x_p,
\eea
where $\delta x_p$ denotes the departure of $x_p$ from its equilibrium value.
The time derivative of $\mu_{\Delta}$ is : 
\bea
\frac{d\mu_{\Delta}}{dt}={\cal C}\omega \frac{\Delta n_{B}}{\bar n_{B}}cos (\omega t)+{\cal B}\bar n_{B}\frac{dx_p}{dt},
\label{diff_eq}
\eea
where,  ${\cal C}$ is defined as  the beta-off-equilibrium baryon density susceptibility and ${\cal B}$ is the beta-off-equilibrium proton fraction susceptibility :  
\begin{align}
{\cal C}\equiv\bar{n}_{B}\left.\frac{\partial\mu_\Delta}{\partial n_{B}}\right|_{x_p}
\quad,\quad 
{\cal B} \equiv\frac{1}{\bar{n}_{B}}\left.\frac{\partial\mu_\Delta}{\partial x_p}\right|_{n_B}\label{eq:susceptibilities}.\end{align}
 These two susceptibilities depend on the equation of state (EoS) of the system. To obtain the temperature and amplitude dependence of the bulk viscosity, we formulate the beta equilibration rate in the presence of trapped neutrinos. We define the net equilibration rate for the relevant processes as follows, 
 \begin{equation}
\Gamma^{\leftrightarrow} 
 \equiv \Gamma^{\rightarrow} -\Gamma^{\leftarrow}
 = \bar n_{B}\frac{d x_p}{d t},
 \label{int_rate2}
\end{equation}
where, $\Gamma^{\rightarrow}$ is the forward interaction rate and $\Gamma^{\leftarrow}$ is the backward interaction rate. By introducing dimensionless variables $\phi\equiv\omega t$, and $A(\phi)\equiv\mu_{\Delta}/T$, we can express Eq.\ref{diff_eq} as follows,
\bea
\frac{dA(\phi)}{d\phi}=d\cos(\phi)+ f \, ,
\label{mu_delta}
\eea
where the prefactors are given by, 
\bea
d\equiv \frac{{\cal C}}{T}\frac{\Delta n_{B}}{\bar n_{B}}, \qquad f\equiv \frac{{\cal B} \Gamma^{\leftrightarrow}}{\omega T}.
\label{prefac}
\eea
Once the function $A(\phi)$ is obtained by solving Eq. \ref{mu_delta}, we can proceed to calculate the bulk viscosity. Bulk viscosity is the response of the system when the system is under repetitive oscillation of compression and rarefaction.  Because of this cyclic process, energy gets dissipated. The energy dissipation rate per volume due to oscillation is given by :  
\begin{equation}
\frac{d\epsilon}{dt}=-\zeta\left(\vec{\nabla}\cdot\vec{v}\right)^{2},\end{equation}
where,  $\vec v$ is the local velocity of the fluid, $\zeta$ is the bulk viscosity. The continuity equation of the conserved number density is given by,
\begin{equation}
    \frac{\partial n_{B}}{\partial t}+\vec{\nabla}\cdot\left(n_{B}\vec{v}\right)=0.
\end{equation}
 Neglecting density gradient ($\nabla n_{B}/\bar{n}_{B}\ll1$) and averaging over an oscillating period one obtains,
\begin{equation}
\zeta\approx-\frac{2}{\omega^{2}}\left\langle \frac{d\epsilon}{dt}\right\rangle \frac{\bar{n}_{B}^{2}}{\left(\Delta n_{B}\right)^2}\,. 
\label{eq:zetadef}
\end{equation}
 Volumetric change of fluid element in density oscillation  is related to the fluctuations of conserved quantity through the relation, $dn_{B}/n_{B}=-dV/V$. The mechanical work done due to the change in volume on the other hand  is described by the expression $d\epsilon=-pdV/V$. The time-averaged $d\epsilon/dt$ can be calculated from the induced pressure oscillation by evaluating following integral,
\begin{equation}
\left\langle \frac{d\epsilon}{dt}\right\rangle=\frac{1}{\tau}\int_{0}^{\tau}\frac{p}{n_{B}}\frac{dn_{B}}{dt}dt,
\label{eq:dedt}
\end{equation}
where, $\tau$ is the oscillation period. The density oscillation leads to variations in pressure, which can be expressed as,  $p=\bar{p}+ \left(\partial p/\partial n_{B}\right)|_{x_p} \delta n_{B}+ \left(\partial p/\partial x_p\right)|_{n_B} \delta x_p$  
 considering small amplitude oscillations $\Delta n_{B}/\bar n_{B}\ll 1$. The equilibrium value $\bar p$ and the second term due to density oscillation does not contribute in the integral written in Eq.(\ref{eq:dedt}).
 The third term can be expressed as:
 $\l(\partial p/\partial x_p\r)|_{n_B}
  = \bar{n}_{B}^{2}\l(\partial \mu_\Delta/\partial n_{B}\r)|_{x_p}$.   The proton fraction changes with time for the weak interaction processes due to density oscillation, $\delta x_p(t)=\int_{0}^{t}(dx_p/dt^\prime)dt^\prime.$ Combining Eqs.(\ref{eq:zetadef}, \ref{eq:dedt}) and using the pressure expression we obtain the final form of the bulk viscosity as:
 \begin{equation}
\zeta=-\frac{1}{\pi}\frac{\bar{n}_{B}^{3}}{\Delta\! n_{B}}\int_{0}^{\tau}\frac{\partial \mu_\Delta}{\partial n_{B}}\int_{0}^{t}\frac{dx_p}{dt^\prime}dt^{\prime}\cos\left(\omega t\right)dt.
\label{zeta}
\end{equation}

In the dynamic environment of a neutron star merger, the rhythmic compression and rarefaction of matter alter the beta equilibration rate. The relaxation of the proton fraction towards equilibrium occurs  through diverse mechanisms, encompassing DURCA, MURCA processes, and neutrino pair bremsstrahlung involving constituent particles.   The DURCA process functions under specific kinematic constraints related to the Fermi momentum of interacting particles and is only active above a certain threshold density. In this study, we investigate the MURCA process in the presence of trapped neutrinos, specifically in scenarios where the threshold conditions for the DURCA process are not satisfied.
To ensure the MURCA  process operates satisfying both energy and momentum conservation, the presence of a spectator particle is necessary \cite{PhysRevLett.12.413,PhysRev.140.B1452,1973ApJ...180..911F,1979ApJ...232..541F}. Below, we outline the two MURCA processes,
\bea
n+N\leftrightarrow N+p+e+\bar\nu_e, \qquad N+p+e\leftrightarrow N+n+\nu_e,
\label{murca}
\eea

where $N$ acts as a spectator particle. 

 Re-equilibration rate is defined to be $\Gamma^{\leftrightarrow}\equiv \Gamma^{\rightarrow}-\Gamma^{\leftarrow}$, where, $ \Gamma^{\rightarrow}$ is the rate of the forward and $\Gamma^{\leftarrow}$ is the backward reaction. Let us denote the rate of the above two processes as $\Gamma_1^{\leftrightarrow}(\Gamma_1^{\rightarrow}(N+p+e+\bar\nu_e\rightarrow n+N)-\Gamma_1^{\leftarrow}(n+N\rightarrow N+p+e+\bar\nu_e)))$ and 
 $\Gamma_2^{\leftrightarrow}(\Gamma_2^{\rightarrow}(N+p+e\rightarrow N+n+\nu_e)-\Gamma_2^{\leftarrow} (N+n+\nu_e\rightarrow N+p+e))$, respectively \cite{2001PhRvD..64h4003J}. The equilibration rate of the first process can be calculated as:
 \bea
\Gamma_1^{\leftrightarrow}=\Gamma_1^{\rightarrow}-\Gamma_1^{\leftarrow}&=&
\int \frac{d^3p_n}{(2\pi)^3}\frac{d^3p_N}{(2\pi)^3}\frac{d^3p_N'}{(2\pi)^3}\frac{d^3p_p}{(2\pi)^3}
\frac{d^3p_e}{(2\pi)^3}
\frac{d^3p_{\nu_e}}{(2\pi)^3}\nn\\
&&(2\pi)^4|M_{fi}|^2
[\delta^4(p_n+p_N-p_N'-p_p-p_e-p_{\bar \nu_e})]{\cal P}_1,
\label{eq:int_rate}
\eea
where, phase space factor ${\cal P}_1$ is given by,
\bea
{\cal P}_1=-\l[f_nf_N\l(1-f_N\r)\l(1-f_p\r)\l(1-f_{\bar\nu_e}\r)\l(1-f_e\r)-f_Nf_pf_ef_{\bar\nu_e}\l(1-f_N\r)\l(1-f_n\r)\r].
\eea
$f_n$, $f_N$, $f_p$, $f_e$, $f_{\nu_e}$ are the distribution functions for neutrons, spectator neutrons, protons, electrons and neutrinos respectively given by, $f_i=(1+e^{\beta (E_i-\mu_i)})^{-1}$, where, $i = n, N, p, e, \nu_e$,  $\mu_i$ denote the chemical potentials for different particles and $\beta=1/k_BT$ ($k_B$ Boltzmann constant).   The antineutrino distribution function is given by $f_{\bar \nu_e}=(1+e^{\beta (E_{\bar\nu_e}+\mu_{\nu_e})})^{-1}$. 

For subsequent calculation, the squared scattering matrix element $|M_{fi}|^2$ in the Eq. (\ref{eq:int_rate}) is given by \cite{doi:10.1146/annurev.astro.42.053102.134013}, 
\bea
|M_{fi}|^2=16G^2\frac{21}{4}\left(\frac{f}{m_{\pi}}\right)^4\frac{g_A^2}{E_e^2}\frac{p_{fn}^4}{(p_{fn}^2+m_{\pi}^2)^2},
\eea
where, $G=8.74\times 10^{-5}$ MeV fm$^3$ 
($1.439 \times 10^{-49}$erg cm$^3$) is the weak Fermi coupling, $g_A=1.26$ is the
 axial vector renormalization, $f\sim1$ is the p-wave $\pi N$ coupling constant in the one pion exchange theory of NN interaction. $p_{fn}$ is the Fermi momentum of neutrons and $m_{\pi}$ is the mass of pion. We consider matrix amplitude to be independent of momentum and energy  and hence can be taken out from the integration. 
 Following  detailed derivation of multidimensional energy and momentum integral in the Appendix.(\ref{int_rate}) one obtains the final form of $\Gamma_1^{\leftrightarrow}$ as,
\bea
\Gamma_1^{\leftrightarrow}&\simeq &\tilde \Gamma T^7
\int dx_{\nu_e}x_{\nu_e}^2\frac{1}{1+e^{(-x_{\nu_e}+\mu_{\Delta}/T)}}\frac{1}{4!}\l[\l(\frac{\mu_{\Delta}}{T}-x_{\nu_e}\r)^4+10\pi^2\l(\frac{\mu_{\Delta}}{T}-x_{\nu_e}\r)^2+9\pi^4\r]
,
\label{gamma1}
\eea
where, $\tilde{\Gamma}=-4.68\times 10^{-19.0}\times(\frac{x_p n_B}{n_0})^\frac{1}{3}\times (\frac{m^{\star } }{m})^4$ MeV$^{-3}$  and other variables are defined in the Appendix.(\ref{int_rate}). In the above expression, the contribution from the antineutrino distribution function is exponentially suppressed under degenerate conditions ($\mu_{\nu_e} \gg T$, $\mu_{\nu_e}$ is chemical potential of neutrino) and is therefore neglected.

Similarly, $\Gamma_2^{\leftrightarrow}$ is expressed in the form shown below, 
\bea
\Gamma_2^{\leftrightarrow}&=&\tilde \Gamma T^7
\int dx_{\nu_e}x_{\nu_e}^2 \l(\frac{1}{1+e^{(-x_{\nu_e}-\mu_{\Delta}/T)}}\r)\frac{1}{4!}\l[\l(\frac{\mu_{\Delta}}{T}+x_{\nu_e}\r)^4+10\pi^2\l(\frac{\mu_{\Delta}}{T}+x_{\nu_e}\r)^2+9\pi^4\r]
\nn\\
&&\l[\l(1-\frac{1}{1+e^{(x_{\nu_e}-\mu_{\nu_e}/T)}}\r)
-\frac{1}{1+e^{(x_{\nu_e}-\mu_{\nu_e}/T)}}\r].
\eea
  The final expression for the MURCA interaction rate ($\Gamma^{\leftrightarrow}$) between baryons and leptons involving both the processes ($\Gamma_1^{\leftrightarrow}+\Gamma_2^{\leftrightarrow}$) thus becomes :
\bea
 \Gamma^{\leftrightarrow}&=&\tilde \Gamma T^7
 \int_0^{\infty} dx_{\nu_e}x_{\nu_e}^2\frac{1}{1+e^{\l(-x_{\nu_e}+\mu_{\Delta}/T\r)}}\frac{1}{4!}\l[\l(A-x_{\nu_e}\r)^4+10\pi^2\l(A-x_{\nu_e}\r)^2+9\pi^4\r]\nn\\
 &+&\l(1-\frac{1}{1+e^{\l(x_{\nu_e}-\mu_{\nu_e}/T\r)}}\r)\l(\frac{1}{1+e^{\l(-x_{\nu_e}-\mu_{\Delta}/T\r)}}\r)\frac{1}{4!}\l[( A+x_{\nu_e})^4+10\pi^2( A+x_{\nu_e})^2+9\pi^4)\r]\nn\\
 &-&\l(\frac{1}{1+e^{\l(x_{\nu_e}-\mu_{\nu_e}/T\r)}}\r)\l(\frac{1}{1+e^{\l(-x_{\nu_e}-\mu_{\Delta}/T\r)}}\r)\frac{1}{4!}\l[\l( A+x_{\nu_e}\r)^4+10\pi^2\l(A+x_{\nu_e})^2+9\pi^4\r)\r].
 \label{MURCA_rate}
 \eea
The above equation depends on  density, temperature of the medium and $\mu_{\Delta}$.
 
 For neutrino transparent matter, with $\mu_{\Delta}/T\ll 1$ the above equation  can be expanded in small powers of $\mu_{\Delta}/T$. After carrying out this expansion and performing energy integrations for various constituent particles, the resulting analytical expression for the interaction rate in neutrino-transparent matter takes the following form \cite{Alford:2010gw}:
\bea
\Gamma^{(\leftrightarrow)}&=&- 4.68\times 10^{-19.0}\l(\frac{x_p n_B}{n_0}\r)^\frac{1}{3}\mu_{\Delta}T^{6} \left(1+ \frac{189\mu_{\Delta}^2}{367\pi^2 T^2} +\frac{21\mu_{\Delta}^4}{367\pi^4 T^4}+\frac{3\mu_{\Delta}^6}{1835\pi^6 T^6} +\cdot \right)\rm{MeV}^{4}.\nn\\
\label{eq:modified-rate}
\eea
In this present computation, we perform a direct integration of the MURCA interaction rate (Eq.\ref{MURCA_rate}) instead of performing the aforementioned approximation {\em i.e} $\mu_{\Delta}/T<<1$.  The Eq.(\ref{mu_delta}) then takes the form of an integro-differential equation, as written below:
 \bea
 \frac{d A}{d\phi}&=&d \cos\left(\phi\right)+\frac{{\cal B}}{\omega}\tilde \Gamma T^6\int_0^{\infty} dx_{\nu_e}x_{\nu_e}^2\l(\frac{1}{1+e^{(-x_{\nu_e}+\mu_{\Delta}/T)}}\r)\frac{1}{4!}\l[\l( A-x_{\nu_e}\r)^4+10\pi^2\l( A-x_{\nu_e})^2+9\pi^4\r)\r]\nn\\
 &+&\l(\frac{1}{1+e^{(-x_{\nu_e}+\mu_{\nu_e}/T)}}\r)\l(\frac{1}{1+e^{(-x_{\nu_e}-\mu_{\Delta}/T)}}\r)\frac{1}{4!}\l[\l( A+x_{\nu_e}\r)^4+10\pi^2\l(A+x_{\nu_e}\r)^2+9\pi^4\r]\nn\\
 &-&\l(\frac{1}{1+e^{(x_{\nu_e}-\mu_{\nu_e}/T)}}\r)
\l(\frac{1}{1+e^{(-x_{\nu_e}-\mu_{\Delta}/T)}}\r)
\frac{1}{4!}\l[\l(A+x_{\nu_e}\r)^4+10\pi^2\l(A+x_{\nu_e}\r)^2+9\pi^4\r].
 \label{general-diff-eq}
 \eea

The second term in the integro-differential equation is the feedback term driven by $f$ with non-linear terms of $\mu_{\Delta}$.  $\mu_{\Delta}$ is then obtained after solving the above integro-differential equation. 

From Eq.(\ref{mu_delta}) and Eq.(\ref{zeta}) the final expression for $\zeta$  becomes,
\begin{equation}
\zeta=\frac{\bar{n}_{B}}{\Delta\! n_{B}} \frac{T {\cal C}}{\pi \omega{\cal B}} \int_{0}^{2\pi} A(\phi, d, f) \rm{cos}(\phi)d\phi \, .
\label{eq:general-viscosity}
\end{equation}

The above equation exhibits dependencies on  ${\cal C}$, ${\cal B}$, $d$  $f$, $\omega$ and $\Delta n_{B}/\bar{n}_{B}$. Numerical technique is employed to solve Eq.(\ref{general-diff-eq}) to obtain $\mu_\Delta$ which we describe in the next section. By substituting $\mu_\Delta$ into Eq. (\ref{eq:general-viscosity}), we   obtain $\zeta$ subsequently.

\section{Results and Discussion}

 In this section, we quantify the dissipation caused by bulk viscosity. Bulk viscosity is the response of the medium linked to deviation from beta equilibrium,  hence, determination of bulk viscous dissipation requires both the beta-equilibration  rate and the beta non-equilibration susceptibilities. First, we present the thermodynamics of the underlying medium for computation of the susceptibilities  and then MURCA interaction rate in neutrino trapped baryonic matter. 

\subsection{Variation of chemical potential}

The calculation of bulk viscosity necessitates the information of the underlying EoS of the hadronic medium,  for evaluation of  the susceptibilities ${\cal B}$ and ${\cal C}$. In the current paper, we evaluate the dissipation coefficient considering two zero-temperature relativistic mean-field (RMF) equations of state, NL3 \cite{PhysRevC.55.540}, and DDME2 \cite{Lalazissis:2005de}.   In the DDME2 model, the DURCA density threshold is never reached \cite{Fortin:2016hny}, making MURCA the dominant process. The NL3 model serves as a reference EOS.
NL3 has density-independent meson-nucleon couplings and nonlinear self-couplings, whereas DDME2 does not have any nonlinear self-coupling terms but meson-nucleon couplings are density-dependent \cite{Nandi:2018ami, Nandi:2020luz}. Here we consider the equations of state at zero temperature. It is worth mentioning that the zero-temperature DDME2 EOS is consistent with all the recent nuclear and astrophysical constraints (see Ref. \cite{Thapa:2021ifv} for a detail analysis), including gravitational wave data \cite{Jiang:2019rcw,Biswas:2020puz}, maximum mass pulsar \cite{Romani:2021xmb}, results from the NICER mission \cite{Riley:2019yda, Riley:2021pdl, Miller:2019cac, Miller:2021qha}, and bounds obtained from the chiral effective field theory \cite{Drischler:2020hwi, Drischler:2020yad}.

\begin{figure}[h]
 \includegraphics[width=0.49\textwidth,clip=]{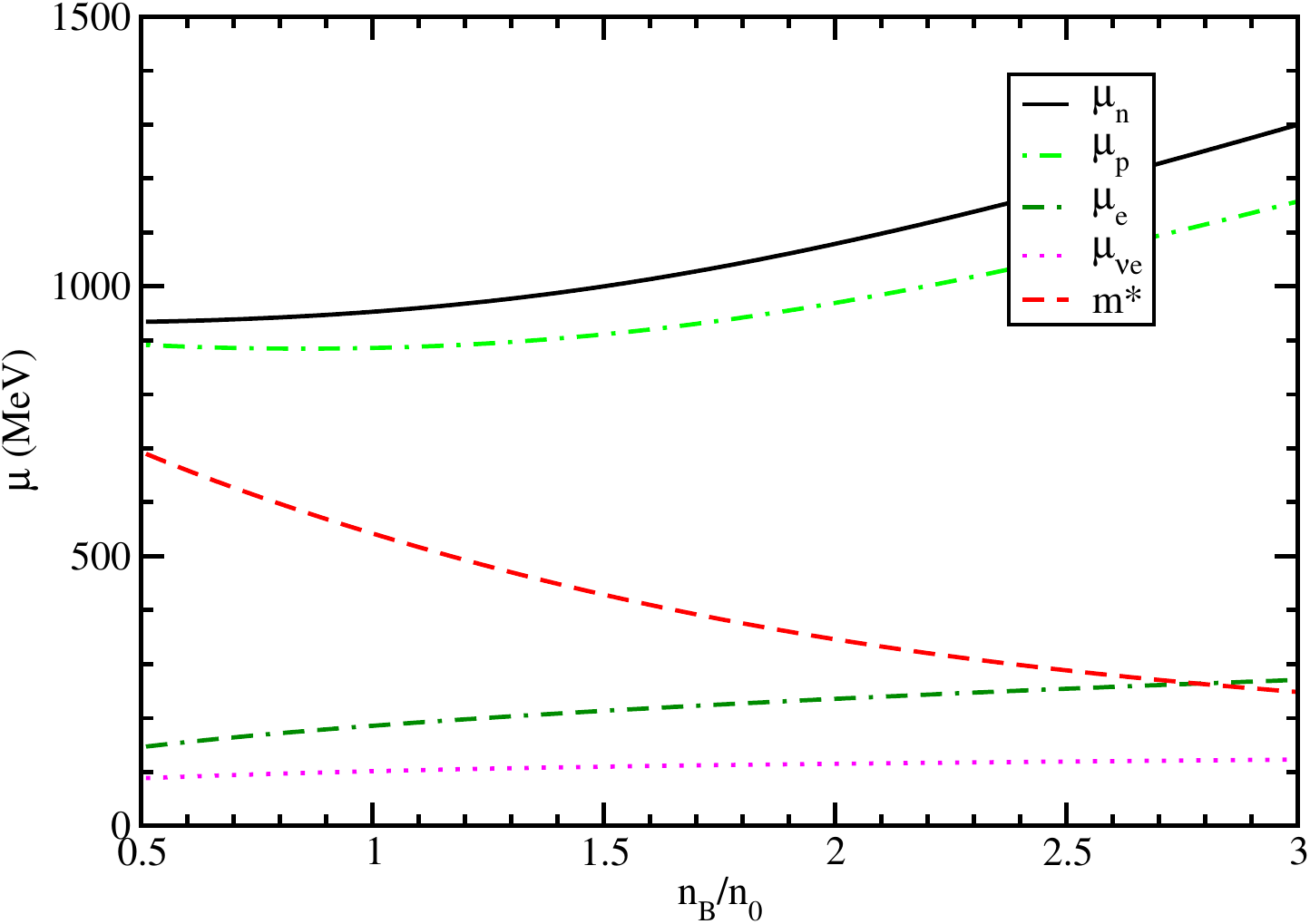}~~~~~~~~~\includegraphics[width=0.49\textwidth,clip=]{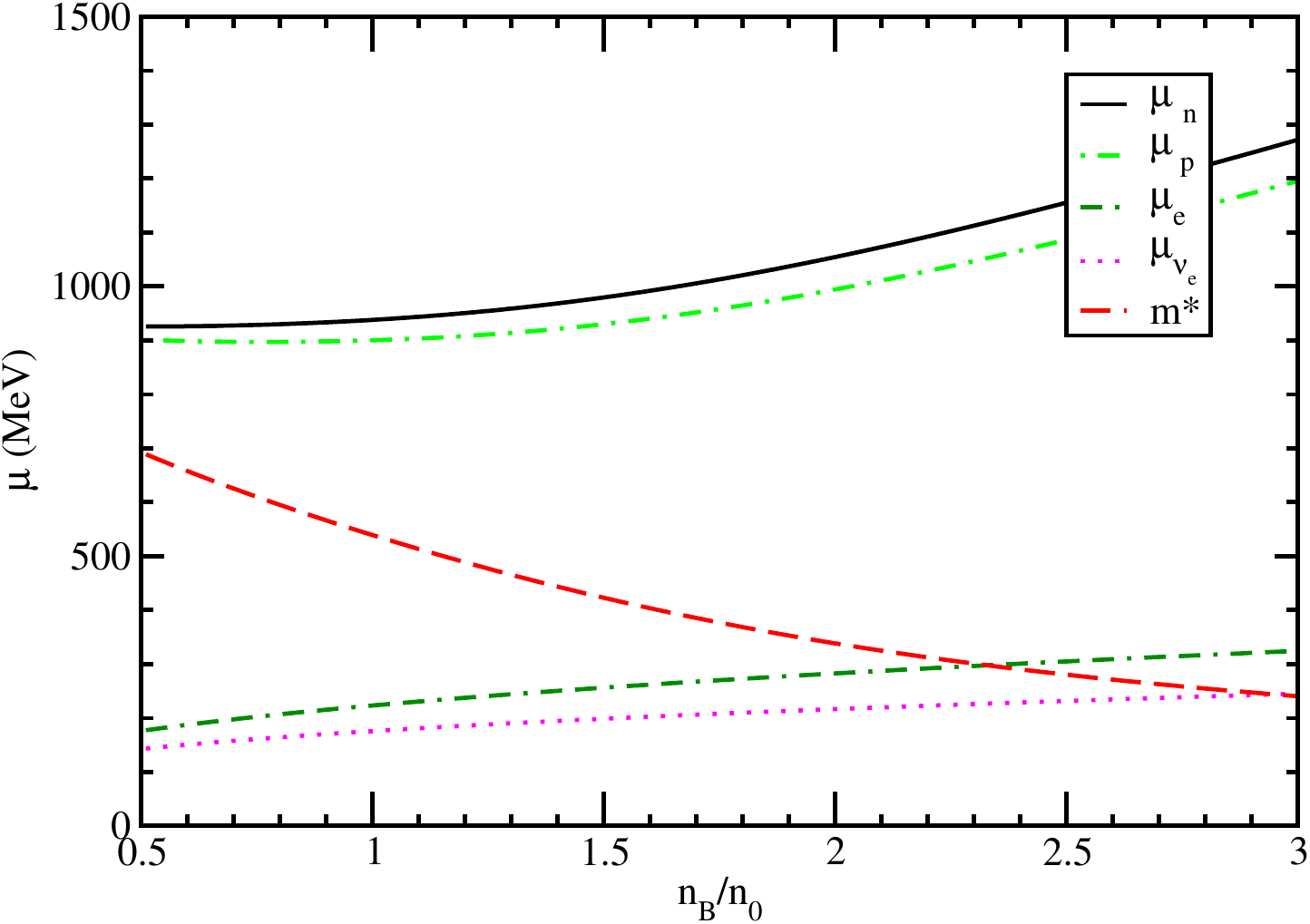}
   \caption{Variation of chemical potential with baryon number density for DDME2 equation of state for $n$, $p$, $e$, $\nu_e$ for $T=0$ and lepton fractions $0.2$ (left) and $0.4$ (right).}
   \label{chempot_DDME2}
 \end{figure}

We obtain medium modified chemical potentials and susceptibilities for both NL3 and DDME2 EOS. The detailed expressions of the chemical potentials as well as susceptibilities  for subsequent numerical analysis are presented in Appendix \ref{EOS}. 
In Fig.(\ref{chem_pot_NL3}), we display the plot illustrating the variation of chemical potentials with density, utilizing the NL3 and the DDME2 EoSs, considering two distinct lepton fractions ($Y_l$).
In the left panel we have plotted chemical potential variation with density for NL3 equation of state for  $Y_l=0.2$ and in the right panel for $Y_l=0.4$.  From the plot it is evident that $\mu_n$ and $\mu_p$ are much higher than $\mu_e$ and $\mu_{\nu_e}$.  In both the plots, we have included curves representing the free chemical potentials ($\mu_{n0}, \mu_{p0}, \mu_{e0}$), i.e. without interactions, for reference.   We also  provide the variation of chemical potential with density for the DDME2 equation of state in Fig. (\ref{chempot_DDME2}) for further comparison. In  Fig.\ref{chempot_DDME2})  the curves for $m^{\star}$    for both the lepton fractions $Y_l=0.2$ and $Y_l=0.4$ have also been presented.

 \begin{figure}[h]
 \includegraphics[width=0.49\textwidth]{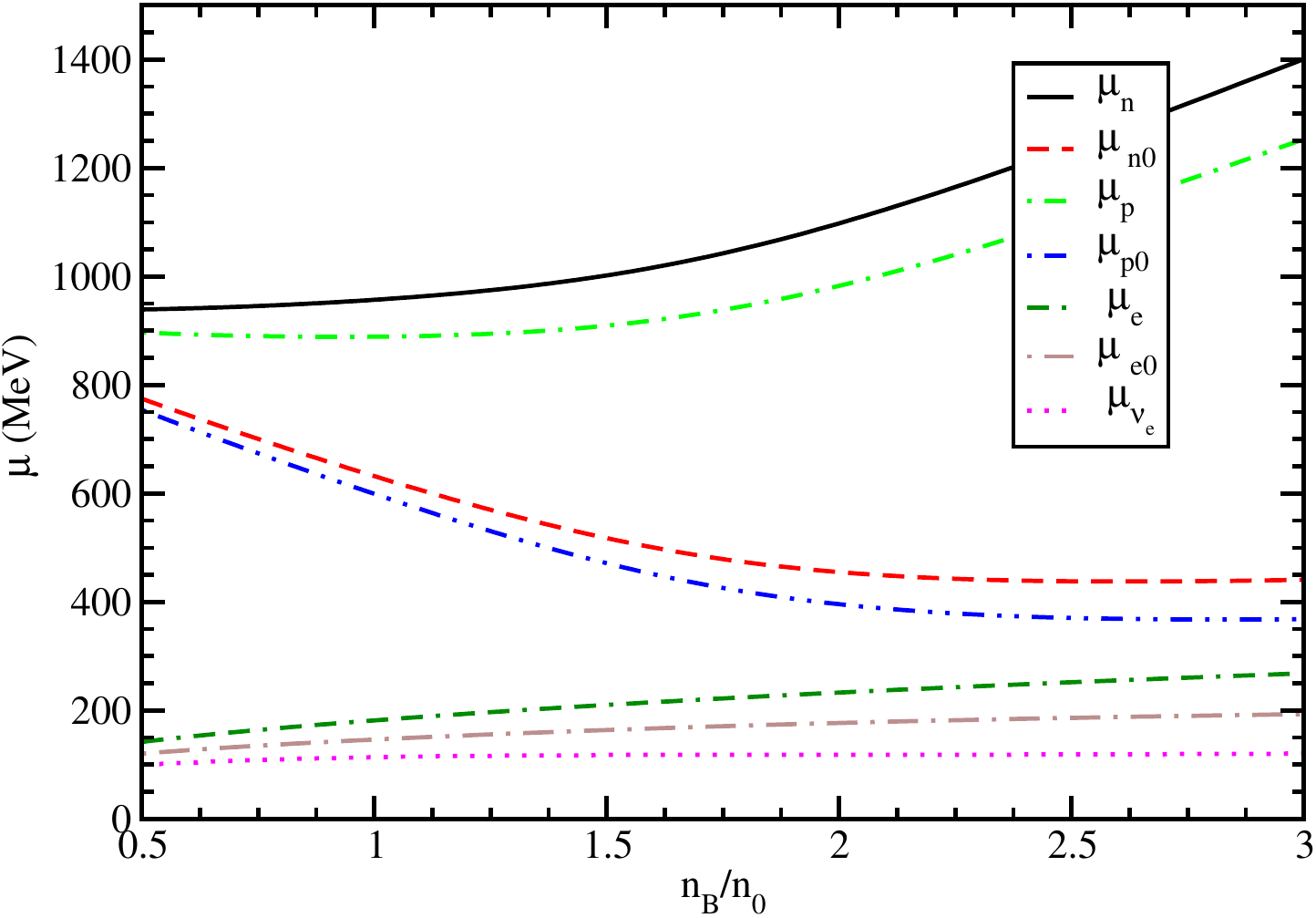}\includegraphics[width=0.49\textwidth]{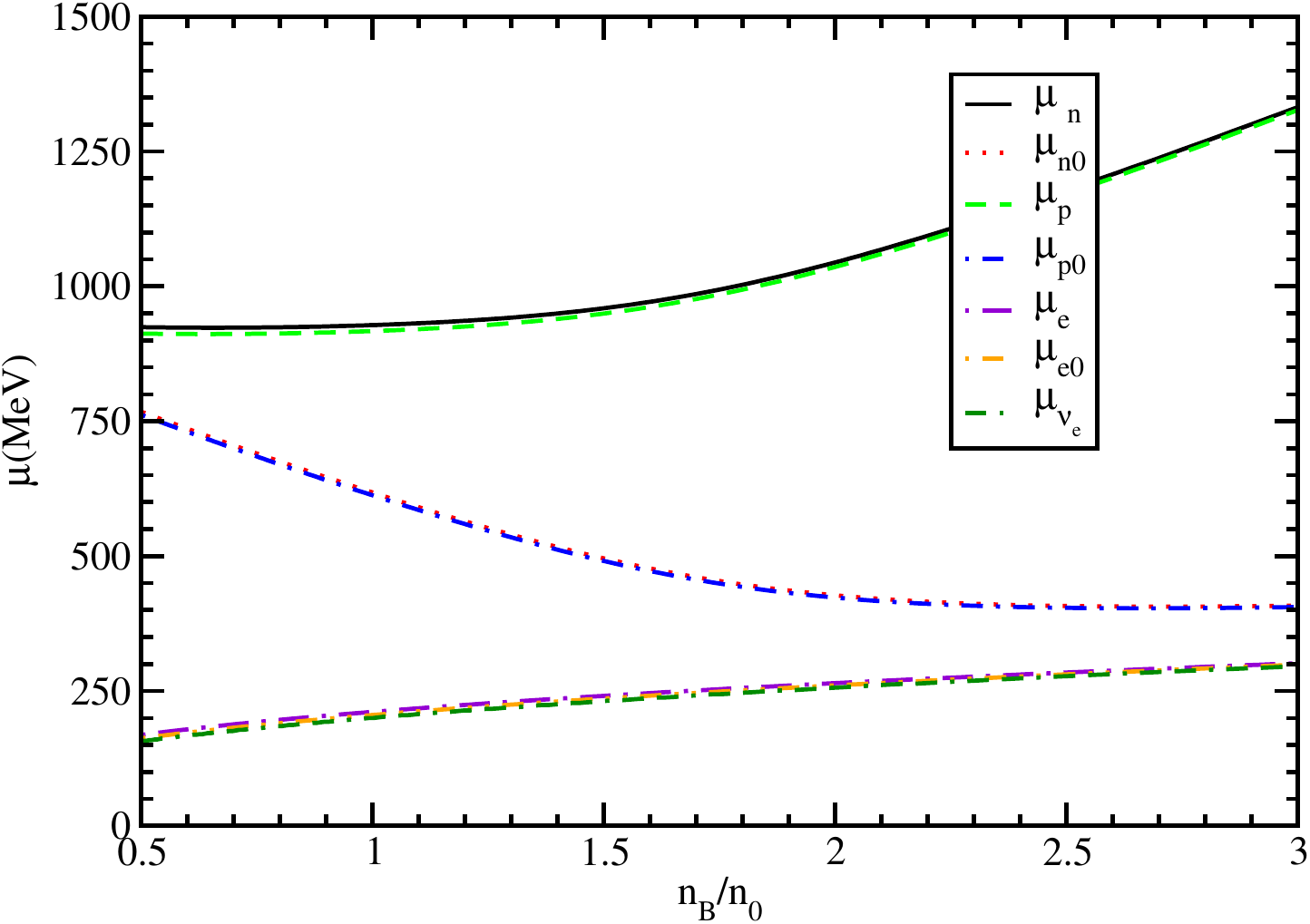}
 
  \caption{Variation of chemical potential with baryon number density for NL3 equation of state for $n$, $p$, $e$, $\nu_e$ at $T=0$ and lepton fractions $0.2$ (left) and $0.4$ (right).}
  \label{chem_pot_NL3}
  \end{figure}

 \subsection{Variation of susceptibility}

\begin{figure}[tbp]
 \includegraphics[width=0.49\textwidth,clip=]{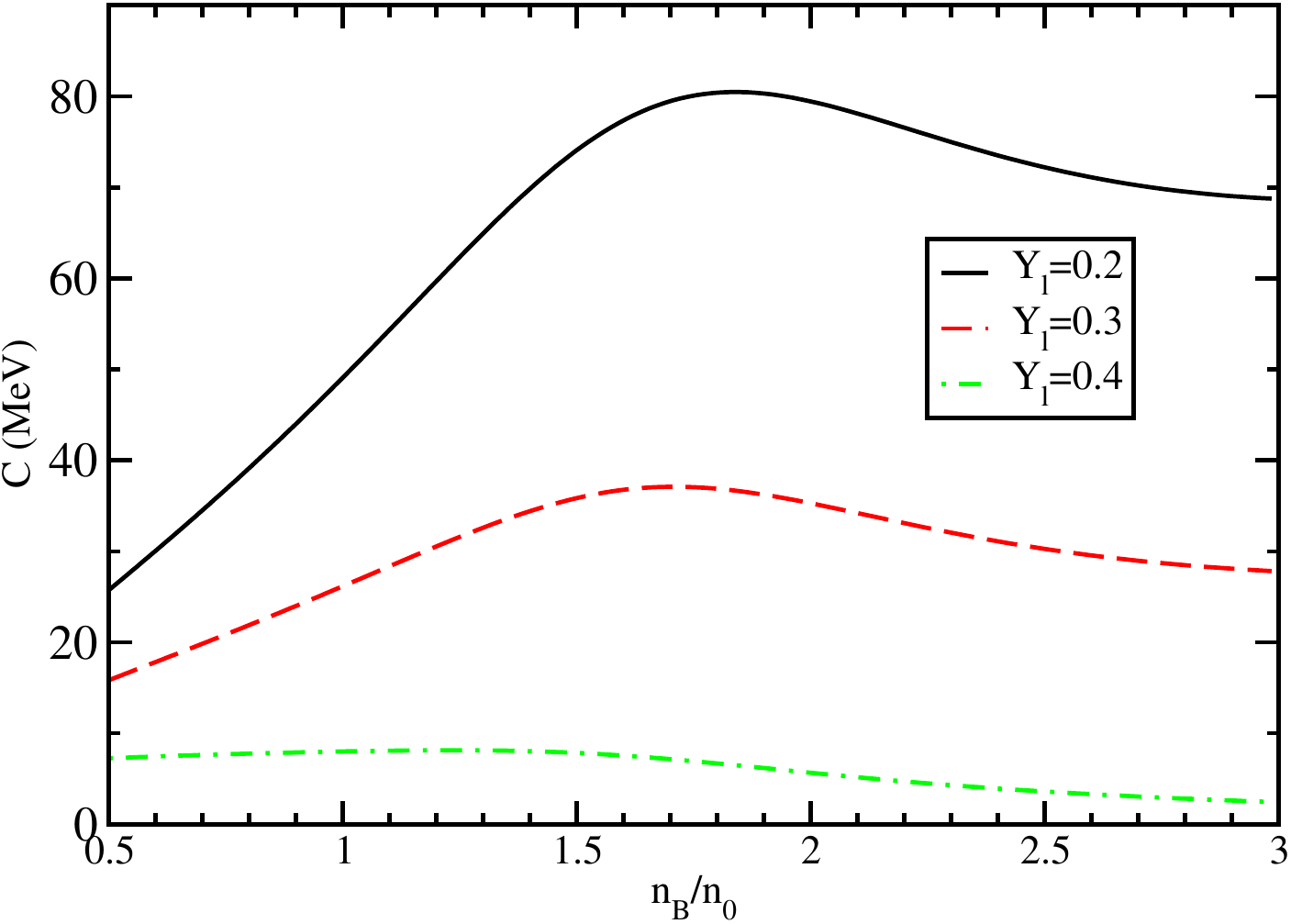}~~~~~~~~\includegraphics[width=0.49\textwidth,clip=]{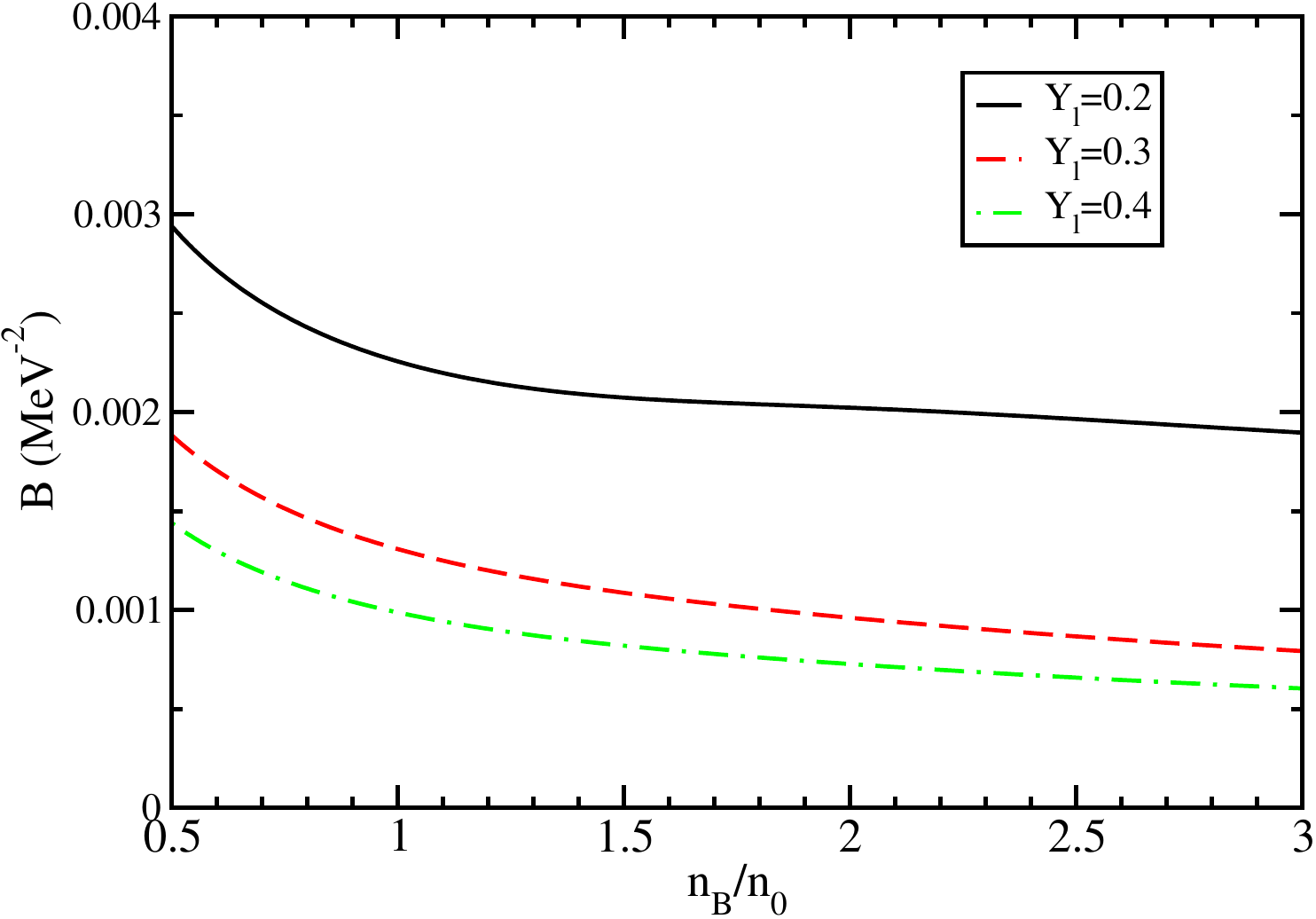}
  \caption{Left: Variation of ${\cal C}$ with baryon number density for different lepton fractions for NL3. Right: Variation of ${\cal B}$ with baryon number density for different lepton fractions for NL3}
  \label{sus_NL3}
 \end{figure}

\begin{figure}[tbp]
 \includegraphics[width=0.49\textwidth,clip=]{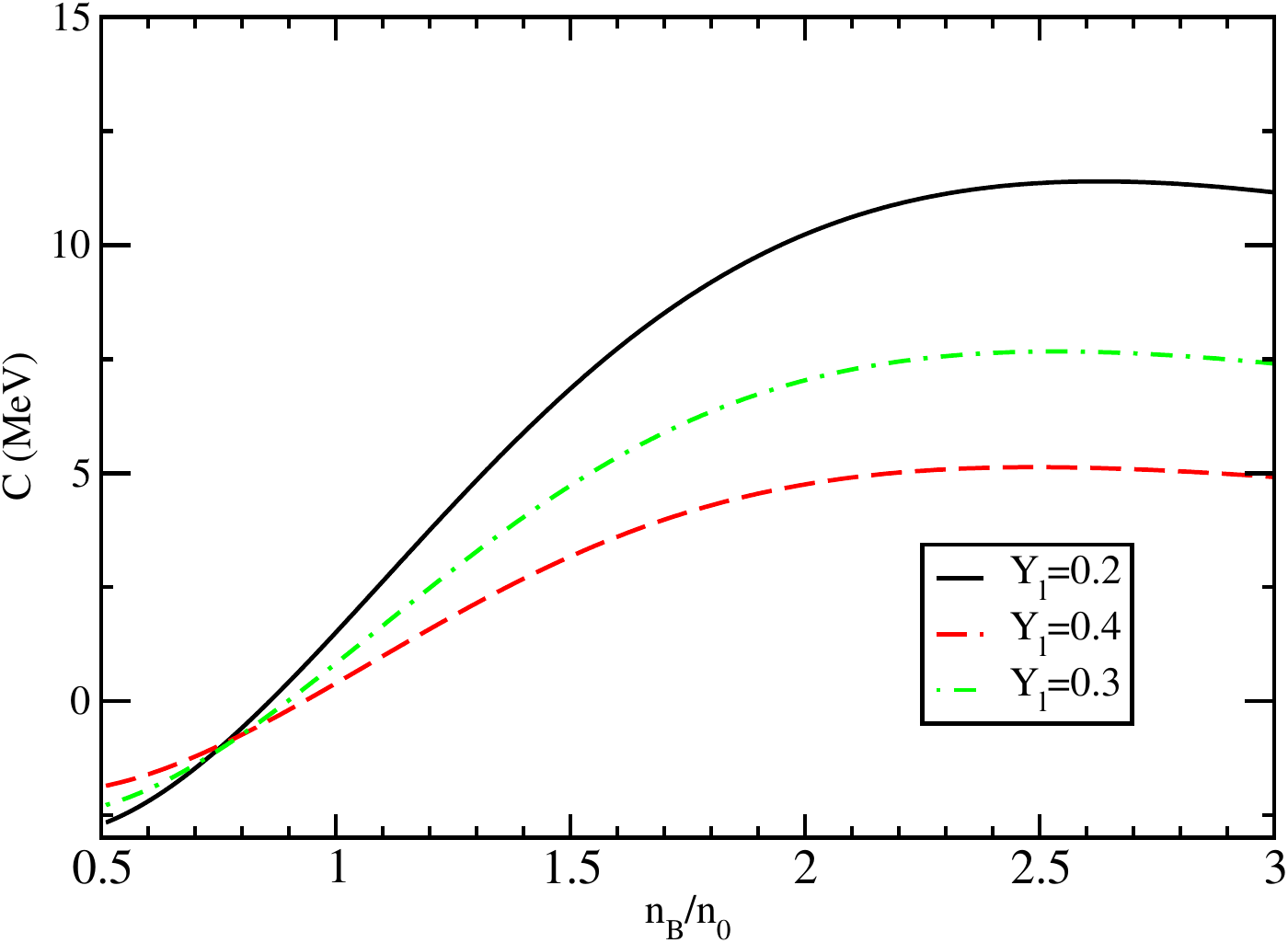}~~~~~~~~\includegraphics[width=0.49\textwidth,clip=]{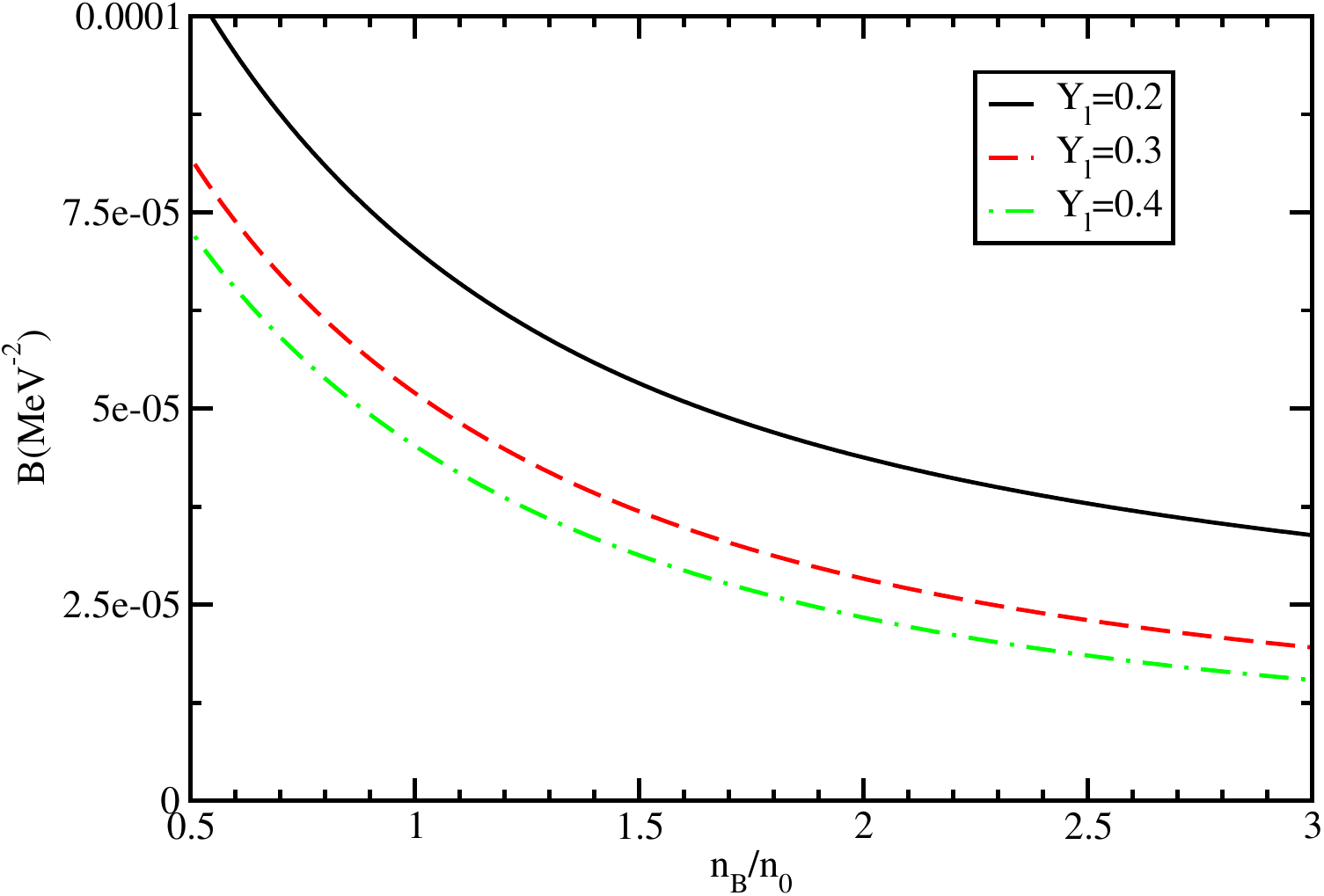}
 \caption{Left: Variation of ${\cal C}$ with baryon number density for different lepton fractions for DDME2. Right: Variation of ${\cal B}$ with baryon number density for different lepton fractions for DDME2}
  \label{sus_DDME2}
 \end{figure}

In Fig. (\ref{sus_NL3}), we illustrate the variations of susceptibilities with baryon density for different lepton fractions. The mathematical expressions for the susceptibilities are given in the 
 Eqn.(\ref{susceptibility}) and Eqn.(\ref{suscep2}). In the left panel, we plot the variation of ${\cal C}$ with $n_B/n_0$ for $Y_l=0.2$, $Y_l=0.3$, and $Y_l=0.4$ for NL3. 
We also include the variation of ${\cal C}$ with number density for the  DDME2 in the left panel of Fig. (\ref{sus_DDME2}). Moving to the right panel of Fig. (\ref{sus_NL3}), we plot the variation of ${\cal B}$ with baryon density for the NL3 equation of state, and in the right panel of Fig. (\ref{sus_DDME2}), we display the same quantity for the DDME2 equation of state. From the plots, we observe that dependence of ${\cal C}$ with density is very prominent. In the lower density range, ${\cal C}$ exhibits an upward trend with density. Beyond a certain critical density, ${\cal C}$ displays weak dependence on baryon density. The threshold density at which this weak dependence occurs is dependent upon the lepton fraction value.   ${\cal B}$ shows less sensitivity to  density variation compared to ${\cal C}$.

In the following subsection we present the variation of bulk viscosity with different parameters like temperature, baryon density. For this first we plot the variation of  $A(\phi)$ with $\phi$ by solving the  integro-differential Eq.(\ref{mu_delta}).  To solve this differential equation, we employ the rk4 algorithm. The energy integration  is performed using the Gauss quadrature technique. Once this integro-differential equation is solved, the obtained $A(\phi)$ is subsequently integrated over $\phi$ to yield the bulk viscosity.
 
 \subsection{Amplitude variation with angular frequency}

In this subsection, we present plots of the general solution of Eq. (\ref{general-diff-eq}), denoted as $ \mu_{\Delta}/T$, as a function of $\phi \equiv \omega t$. The plots are based on the NL3 equation of state. The solution depends on two quantities, namely $d$ and $f$ as already defined in Eq.(\ref{prefac}). For a fixed value of $\Delta n_{B}/\bar n_{B}=10^{-2}$  the values of $d$ and $f$ vary with  temperatures.

\begin{figure}[h]
 \includegraphics[width=0.49\textwidth,clip=]{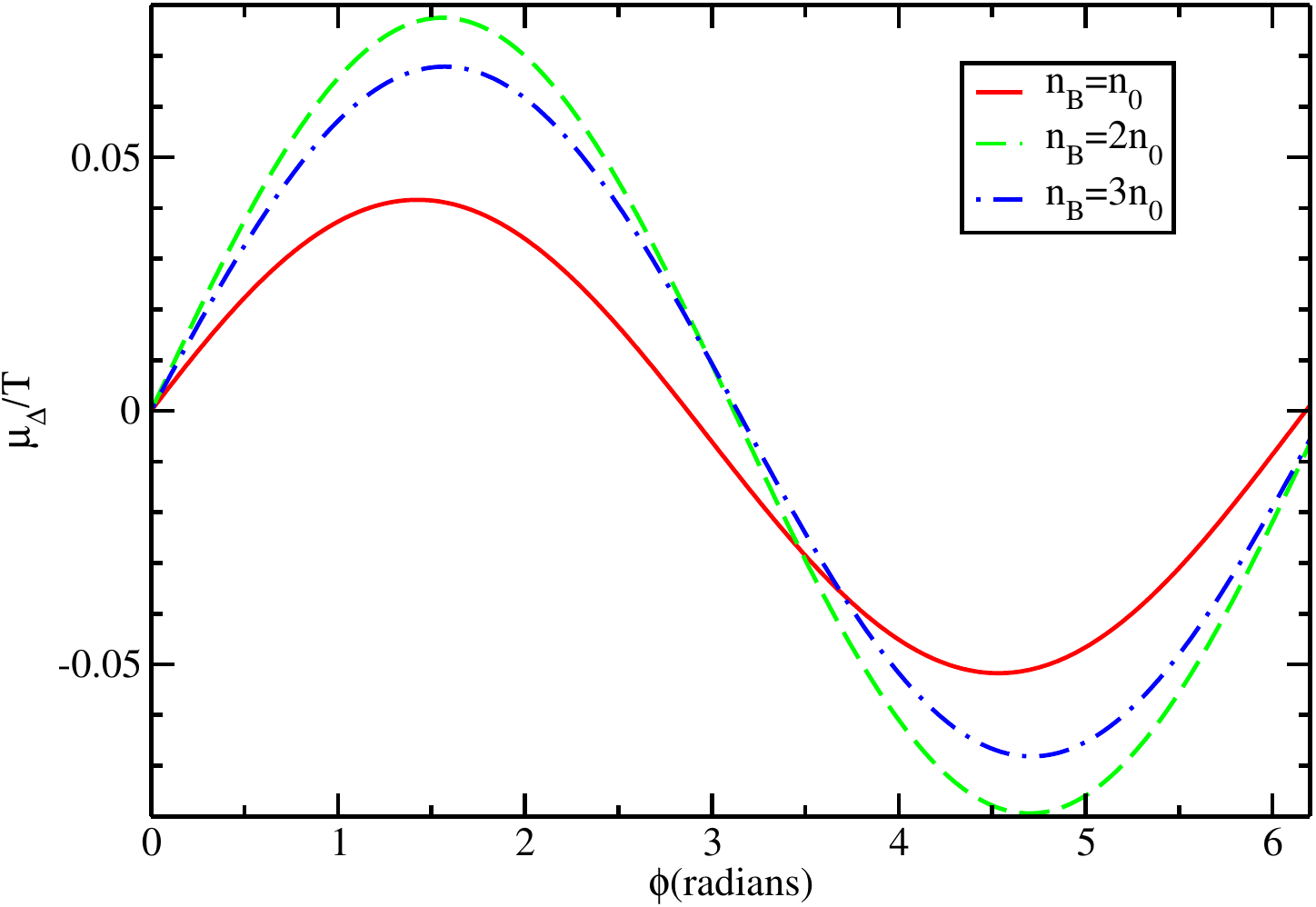}~~~~~~~\includegraphics[width=0.49\textwidth,clip=]{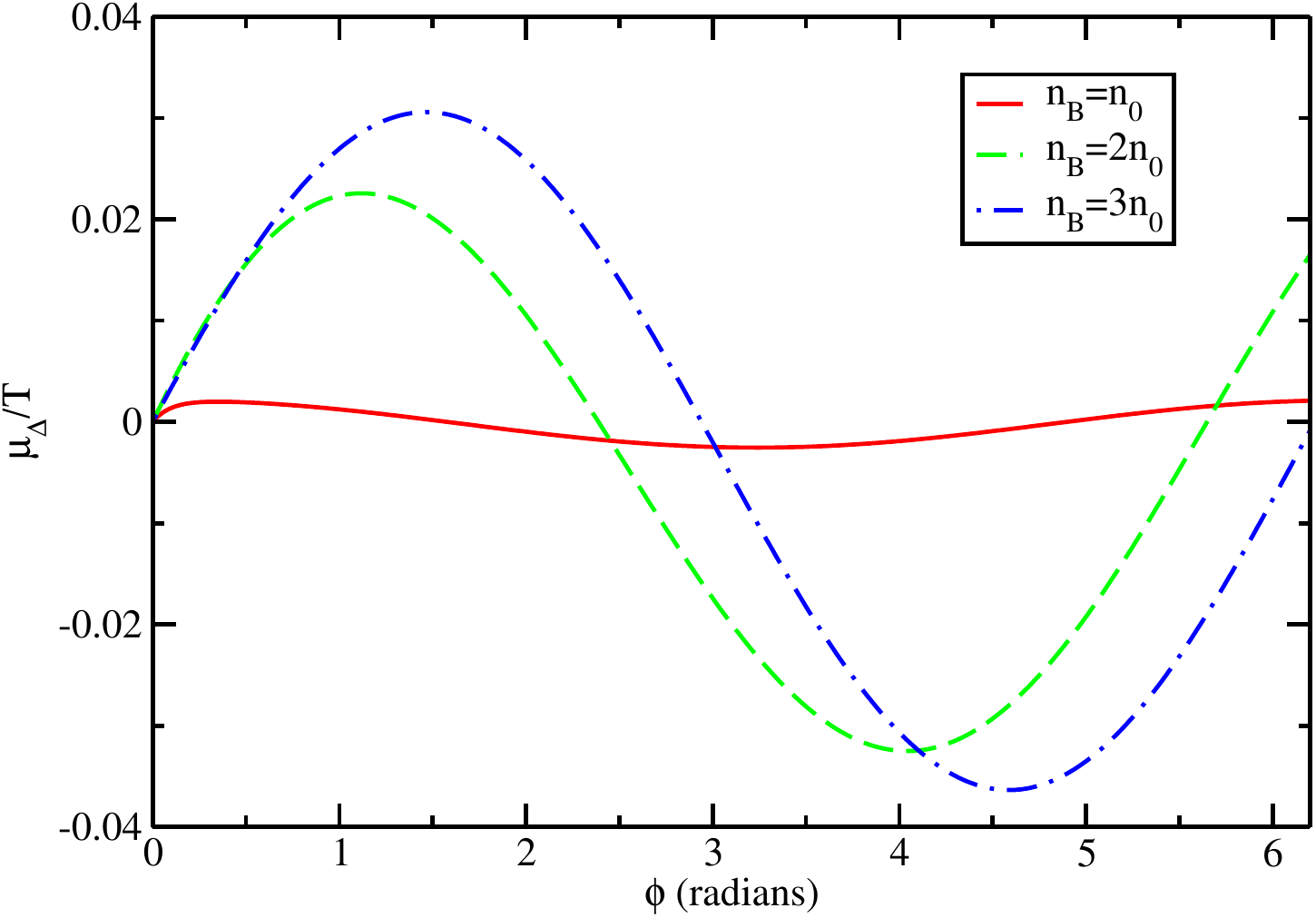}
 
 \caption{Left: Variation of waveform $\mu_{\Delta}/T$ with $\phi$ of particle interaction rate at $T=10$ MeV. Right:  Variation of waveform $\mu_{\Delta}/T$ with $\phi$ of particle interaction rate at $T=20$ MeV for $Y_l=0.2$ using NL3 EoS.}
 \label{mu-delta}
 \end{figure}

In the left plot, we present the curves for different densities, specifically $n_B=n_0$, $n_B=2n_0$, and $n_B=3n_0$, at a temperature of $10$ MeV. On the right, we present the plot of $ A$ for the same densities, but at a higher temperature of $20$ MeV. Increasing density from $n_0$ to $2n_0$ leads to an increment in $A$ and it decreases from $2n_0$ to $3n_0$. This can be explained in this manner, $A$ relies on the susceptibilities ${\cal B}$ and ${\cal C}$ through the parameters $d$ and $f$. As depicted in Fig.(\ref{sus_NL3}), ${\cal C}$ displays a noticeable dependence on density, while ${\cal B}$ exhibits weak dependency. This leads to a substantial density variation in $d$ and a weaker dependency in $f$. Consequently, ${\cal A}$ demonstrates a behavior akin to that of $d$ across different densities. 
Moreover, as the temperature rises, there is a reduction in $d$, leading to a decrease in $ A$, as illustrated in Fig. (\ref{mu-delta}).

\subsection{Variation of bulk viscosity with temperature}
In this subsection, we present the variation of $\zeta$ with temperature. We consider the temperature and density of the hadronic medium to ensure that the semi-degeneracy condition is maintained, i.e., $\mu_i > T$ (where $i=n$, $p$, $e$, $\nu_e$). 
The temperature-density values selected for the calculation adhere to the physical conditions applicable for the merging scenario as well as satisfy the degeneracy condition.

\begin{figure}[tbp]
 \begin{center}
 \includegraphics[width=0.54\textwidth,clip=]{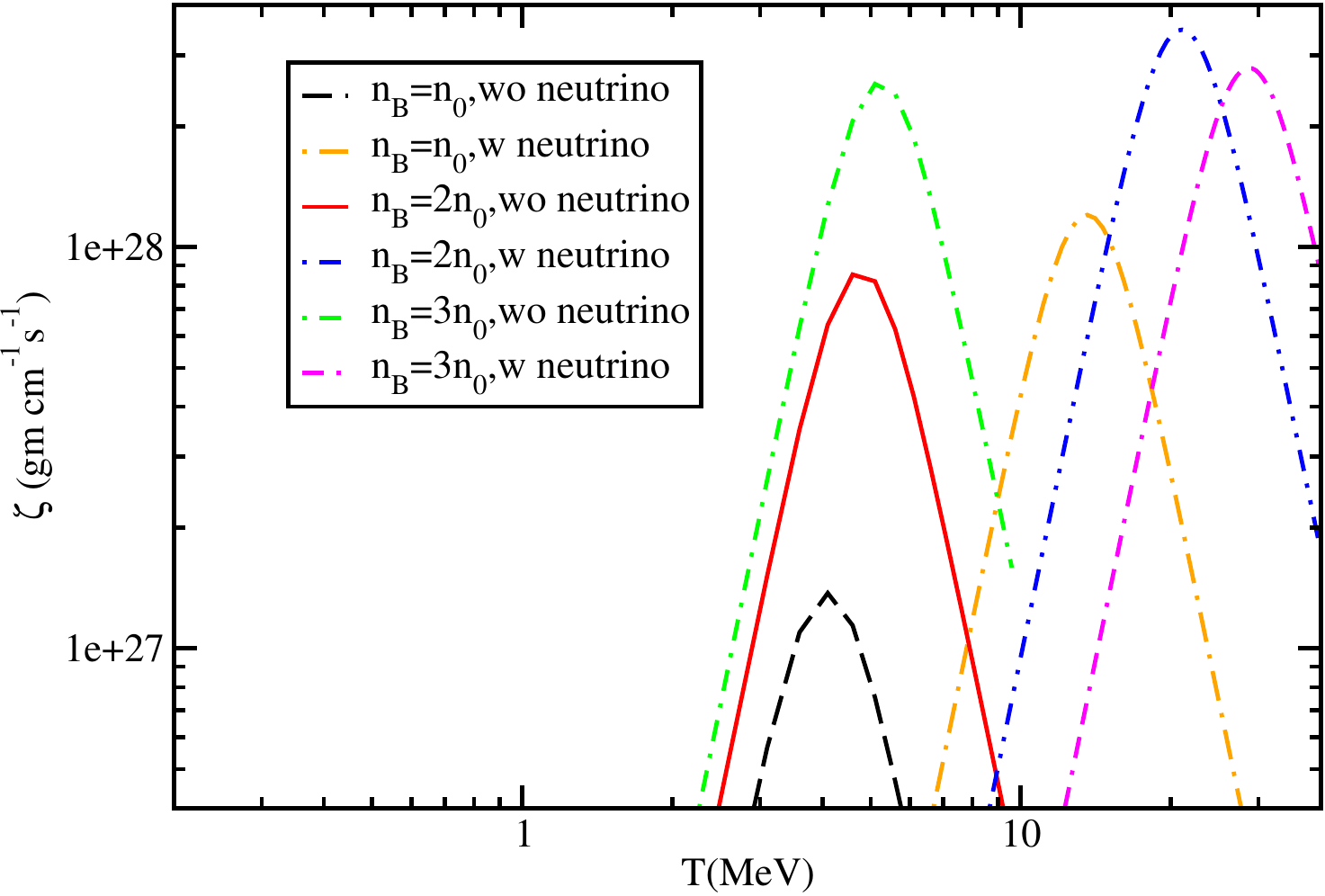}
 \end{center}
  \caption{Variation of $\zeta$  with temperature with and without trapped neutrinos in the interaction rate and in the EoS.}
 \label{fig:neut_plot}
 \end{figure}

In Fig.(\ref{fig:neut_plot}),  we present plots of $\zeta$, illustrating its temperature variation while comparing cases with and without neutrino chemical potential. The black, red and green curves represent the bulk viscosity of baryonic matter without trapped neutrinos. These curves are plotted considering free hadron gas EoS without neutrinos. In this EoS the susceptibilities are given by ${\cal B}=4m_n^2/3(3\pi^2)^{\frac{1}{3}}n_B^{\frac{4}{3}}$ and
${\cal C}=(3\pi^2 n_B)^{2/3}/6m_n$ ($m_n$ is the bare nucleon mass).
Specifically, the black dashed curve corresponds to $n_B=n_0$, the red solid curve corresponds to $n_B=2n_0$ and the green dashed-dotted curve corresponds to $n_B=3n_0$. 
On the other hand, the  orange, blue and magenta curves represent the bulk viscosity of trapped neutrino baryonic matter. These curves are plotted with NL3 EoS for lepton fraction $0.2$. The orange dashed dotted curve corresponds to density $n_B=n_0$, the blue double dot-dashed curve corresponds to density $n_B=2n_0$, and the magenta double dashed-dotted curve corresponds to density $n_B=3n_0$.
From these curves, it is observed that the plot reaches its maximum at a temperature of $4.10$ MeV when the neutrino chemical potential is zero. Employing free EoS, the maximum values of bulk viscosity  for $n_B=n_0$, $n_B=2n_0$, $n_B=3n_0$ are given by $1.45\times 10^{27}$ gm cm$^{-1}$ s$^{-1}$,  $8.62\times 10^{27}$ gm cm$^{-1}$ s$^{-1}$ and $2.57\times 10^{28}$ gm cm$^{-1}$ s$^{-1}$ respectively. 
Considering non-zero $\mu_{\nu_e}$, the maxima of the bulk viscosity shift to $T_{\zeta_{max}}=14.1$ MeV when density is $n_B=n_0$ with an increment of peak value by a factor of $8.25$, for the density $n_B=2n_0$ maxima of the curve is present at  $T_{\zeta_{max}}=21.1$ MeV with an increment in peak value by a factor of $3.23$  and for $n_B=3n_0$ the peak position is at $T_{\zeta_{max}}=28.6$ MeV with an increment in peak value by a factor of $1.33$. 
All these increments are with respect to the neutrino-transparent scenario. The peak position and peak value of the $\zeta$ vs $T$ curve can be  written as $T_{\zeta_{max}}\propto \omega /(\Gamma^{\leftrightarrow} {\cal B})^{1/m}$ ($m>1$) and $\zeta_{max}\propto{\cal C}^2\tau/(4\pi {\cal B})$.
Hence, if $\Gamma^{\leftrightarrow}$,  ${\cal B}$ and ${\cal C}$ change due to incorporation of non-zero $\mu_{\nu_e}$ in the particle interaction rate and in the EoS $\zeta_{max}$ and $T_{\zeta_{max}}$ also change.

\begin{figure}
 \includegraphics[width=0.49\textwidth,clip=]{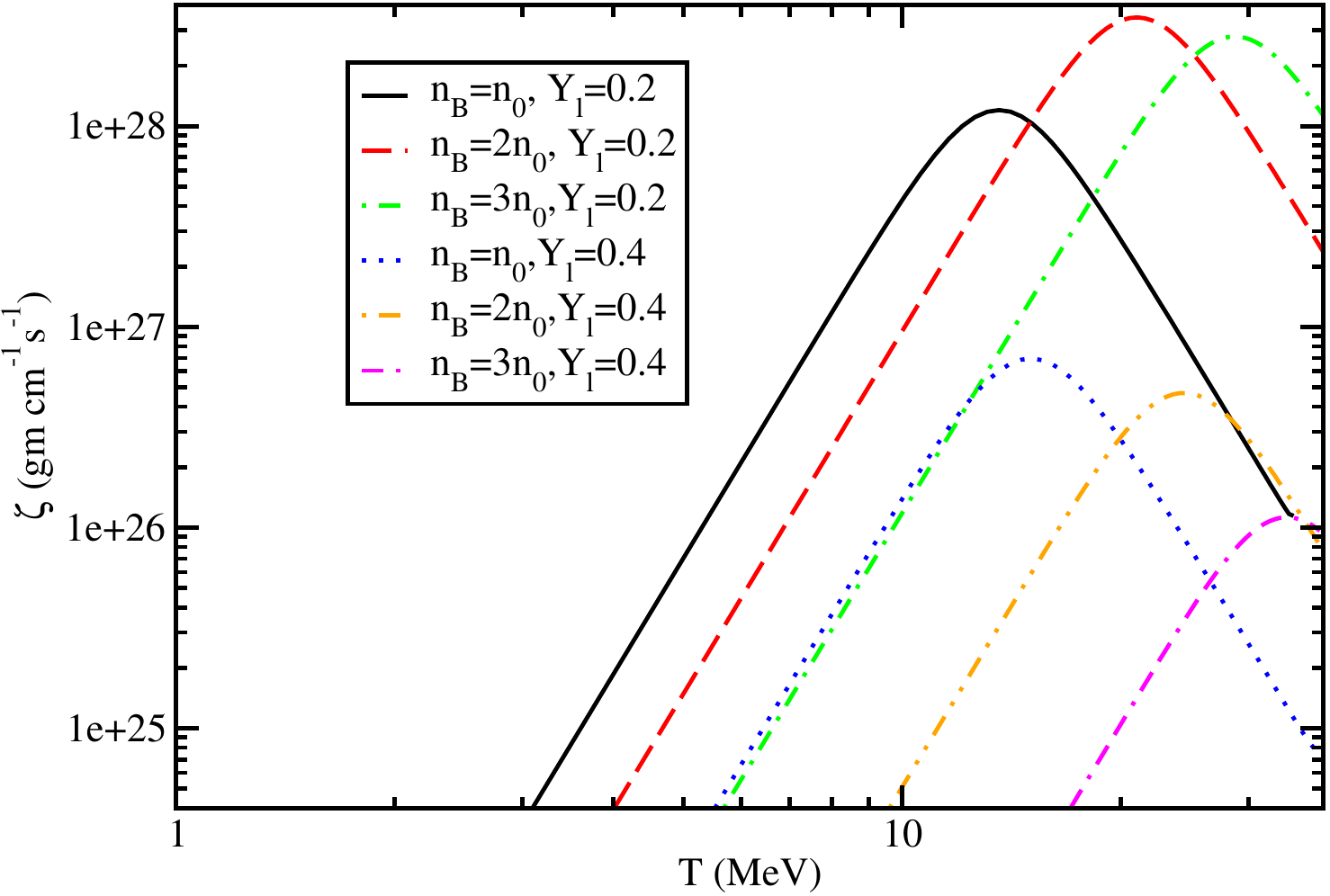}~~~~~~~~~~~~~~~~\includegraphics[width=0.49\textwidth,clip=]{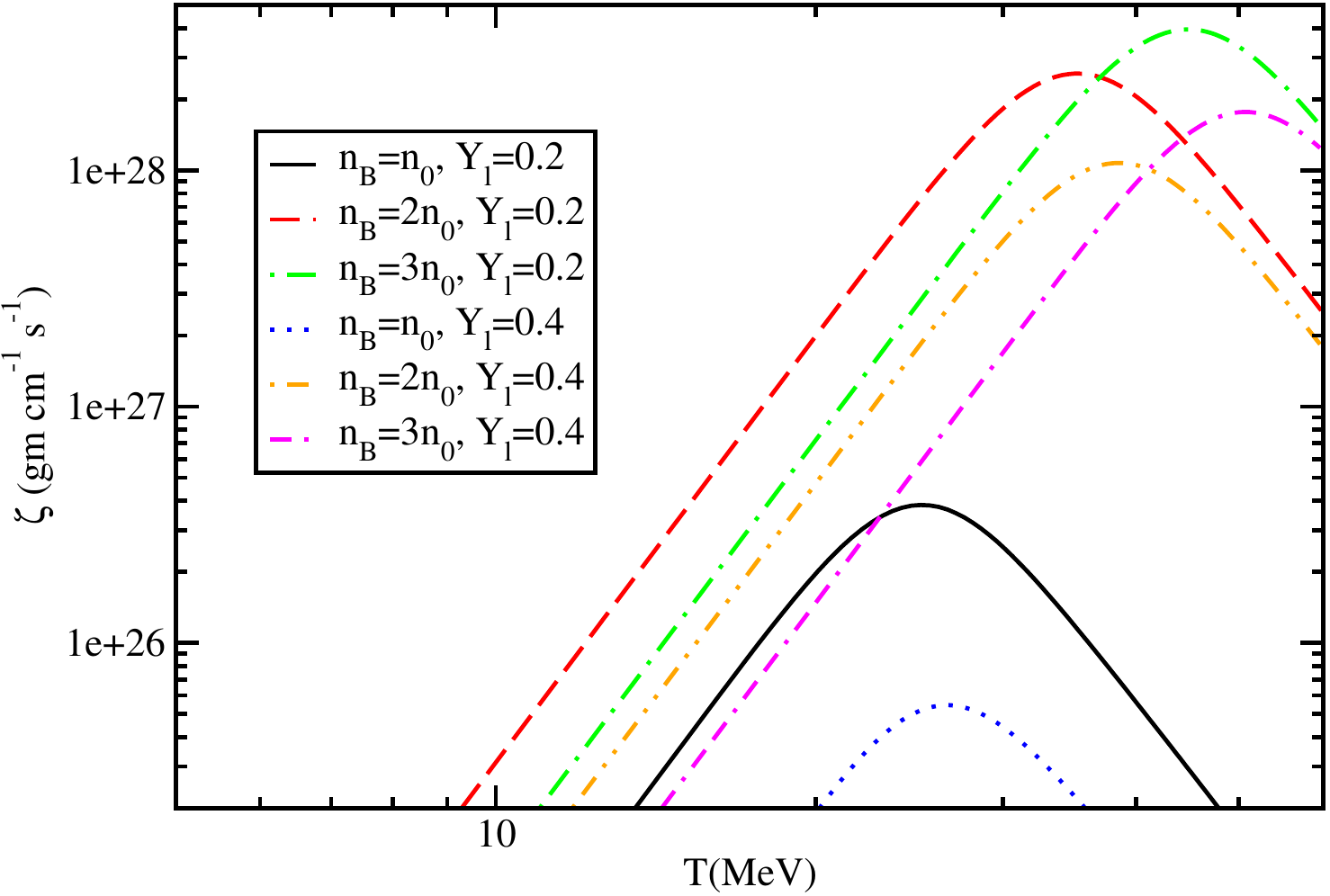}
 \caption{Left: Variation of $\zeta$ with $T$ considering NL3 EoS, Right: Variation of $\zeta$ with $T$ considering DDME2 EoS.}
 \label{vary_temp}
 \end{figure}

Next, in the Fig. (\ref{vary_temp}), we plot the variation of $\zeta$ with temperature for different densities. The left panel considers the NL3 EoS, while the right panel shows the results for the DDME2 EoS. In both the left and right plots, we consider $\zeta$ for $Y_l = 0.2$ and $Y_l = 0.4$. The solid black, red dashed, and green dashed-dotted curves correspond to $n_B=n_0$, $n_B=2n_0$, and $n_B=3n_0$, respectively, for $Y_l=0.2$. The dotted blue, orange double dotted-dashed and magenta double dashed-dotted curves represent $n_B=n_0$, $n_B=2n_0$, and $n_B=3n_0$, respectively, for $Y_l=0.4$. It is observed that the height of the maxima changes with the lepton fraction. Bulk viscosity attains higher maximum values for lower lepton fractions, and this observation applies to both EOSs. This is because higher lepton fraction yields higher interaction rate and hence lower viscosity ($\zeta\propto 1/\Gamma^{\leftrightarrow(m-1)}$).
 
\subsection{Variation with density}

\begin{figure}
 \includegraphics[width=0.79\textwidth,clip=]{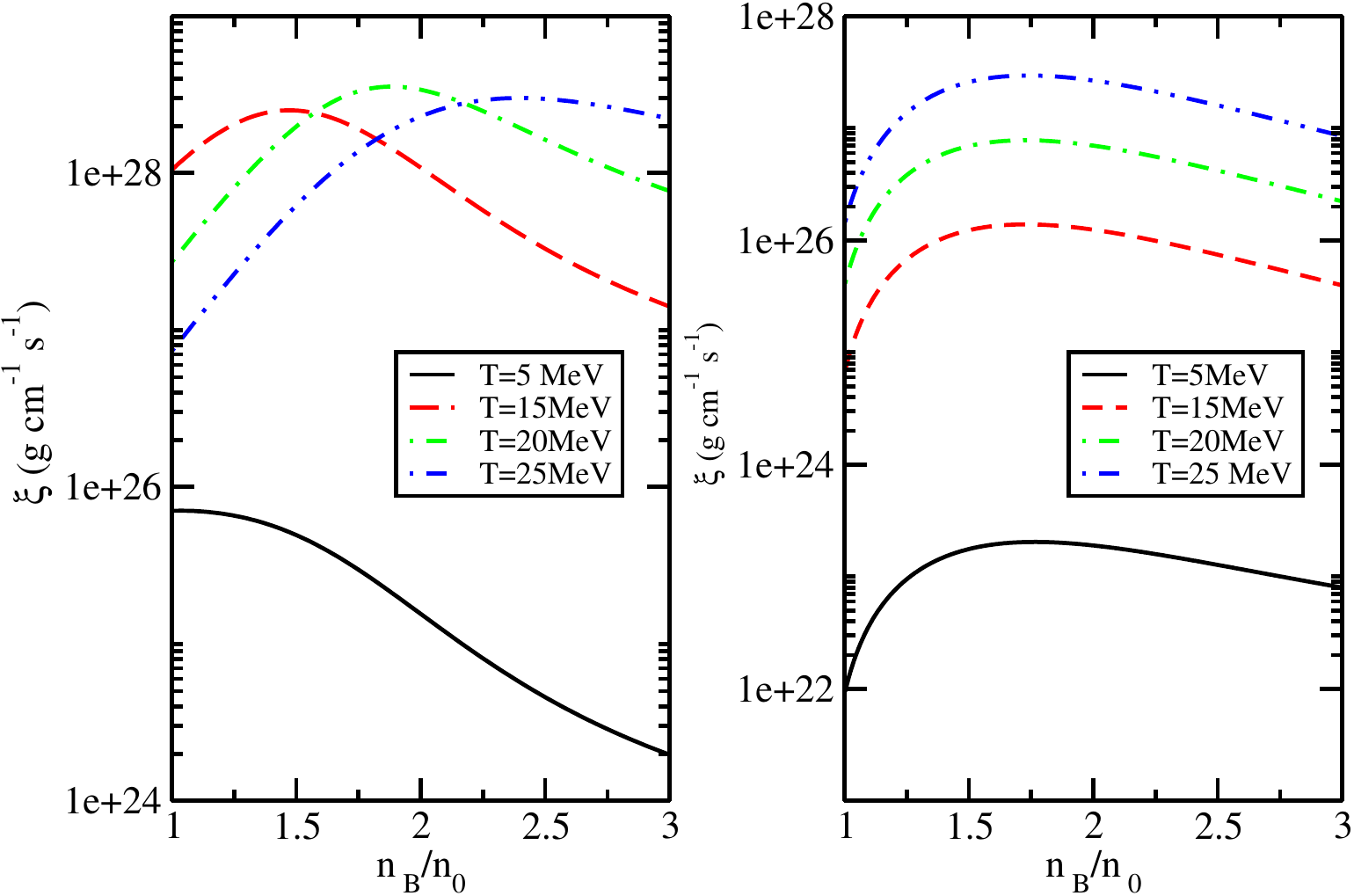}
 \caption{Left: Density variation of $\zeta$ considering NL3 EoS for $Y_l=0.2$. Right: Density variation of $\zeta$ considering DDME2 EoS for $Y_l=0.2$. }
 \label{vary_dens}
 \end{figure}

The Fig.(\ref{vary_dens}) presents the density variation of $\zeta$ for different temperatures. The left panel displays the curves for NL3 EoS, while the right panel shows the corresponding results for DDME2 EoS. In the left panel, $\zeta$ is presented in black solid curve at temperatures $T=5$ MeV, $T=15$ MeV curve is presented in red dashed curve, $T=20$ MeV curve is presented in green dashed-dotted curve and $T=25$ MeV plot is presented in blue double dot-dashed curve. All these curves are  for  lepton fraction $Y_l=0.2$.  In the left panel we plot the curves at same temperatures but for DDME2 EoS.
From the plots, it is evident that the  $\zeta$  first increases and then decreases with density.  The density variation of $\zeta$ is closely linked to the behavior of ${\cal C}$ as can be seen from the Fig.(\ref{sus_NL3}) and only minimally influenced by ${\cal B}$. For DDME2 EoS $\zeta$ is independent of density in the density range $n_B>1.5n_0$.  In all the plots of the Figures (\ref{fig:neut_plot}), (\ref{vary_temp}) and (\ref{vary_dens}), the oscillation frequency is set at $8.4$ kHz.

 \subsection{Estimation of viscous dissipation time scale}
 
 The characteristic time scale of density oscillation, denoted by $\tau_\zeta$, is determined by the ratio of the energy density $\epsilon$ to the dissipated power per unit volume $d\epsilon/dt$. The energy density of baryon number density oscillation is given by $\epsilon = K(\Delta n_B/n_B)^2/18$ and $d\epsilon/dt=\omega^2\zeta(\Delta n_B/n_B)^2/2$.  Substituting this into the expression for $\tau$, we obtain $\tau =Kn_B/(9\omega^2\zeta).$
 Here, $K$ represents the nuclear compressibility of baryonic matter, calculated from the NL3 EOS, as shown in the plot's inset.  The angular frequency of the compact star is considered to be $8.4$ kHz and the temperature is $20$ MeV.  The three-dimensional plot provides the time scale associated with the bulk viscous dissipation coefficient.  From the plot, it can be observed that the timescale varies between approximately $\tau \approx 32\times 10^{-3}$ s to $100\times 10^{-3}$ s in the baryon density range from $n_0$ to $2n_0$. Thus, the timescale aligns with the survival time period of the merged compact object within the mentioned density regime. Beyond $2n_0$, the timescale exceeds the typical survival period of the compact object after merging.

\begin{figure}
\begin{center}
 \includegraphics[width=0.55\textwidth,clip=]{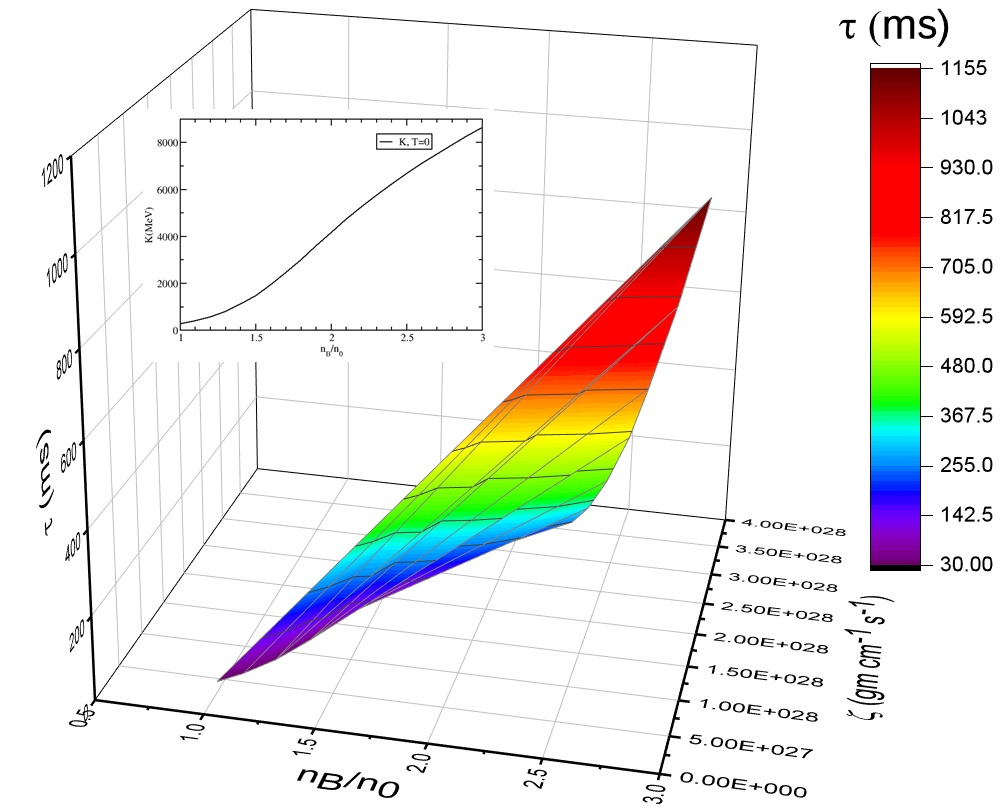}
 \caption{ 3-d variation of $\tau$ with $n_B/n_0$ and $\rm{Log \zeta}$ at T=20 MeV.} 
 \label{time_scale}
 \end{center}
 \end{figure}
 \section{Summary and Conclusion}

In this study, we have  formulated  MURCA driven bulk viscosity   in a nuclear medium consisting of baryons (neutrons and protons) and leptons (electrons and neutrinos). The primary focus of this examination lies in its applicability in the context of binary neutron star mergers.
  In the merging event, the temperature can rise significantly, reaching values as high as $100$ MeV, while the density can reach up to $5n_0$. At these extreme conditions, neutrinos remain trapped within the baryonic matter. Particularly, around a temperature of $T=5$ MeV, the neutrino free path becomes smaller than the radius of the star, resulting in non-zero chemical potential for neutrinos.  For our calculations, we consider  neutrino-trapped baryonic medium at temperature of approximately $T\sim 50$ MeV and a density of around $\sim 3n_0$.

 The current calculation involves two main components: first, preparing the underlying medium with the neutrino-trapped nuclear equation of state, and second, calculating the bulk viscosity by evaluating the neutrino-trapped MURCA  interaction rate. This study incorporates the following distinctive features:

i. Equation of state dependence:

We have considered the NL3 and DDME2 EoSs at zero temperature for our investigation. DDME2 employs a density-dependent parametrization, while NL3 adopts a non-linear parametrization. The medium-modified chemical potentials of the constituent particles have been plotted against density for these two nucleonic EOSs. We have neglected the abundances of anti-baryons in the EoSs since their contributions are insignificant at the temperatures and densities considered in this work. Moreover, for a typical value of $\mu_e=200$ MeV (see Fig. \ref{chem_pot_NL3} and \ref{chempot_DDME2}) and $T=50$ MeV, the weak processes involving positrons will be subdued approximately by a factor $\exp(-\mu_e/T) \simeq 0.02$.  For lower temperatures, it becomes more insignificant. Additionally, we have ignored the direct effect of finite temperature on the EOS, which may be important at $T\gtrsim 20$ MeV. We plan to include all such temperature effects in a future study.
In the present work, therefore, the temperature dependencies only arises through the evaluation of particle interaction rates.

ii. Susceptibility:  
 
 The determination of bulk viscosity necessitates the calculation of susceptibilities ${\cal B}$ and ${\cal C}$. 
 Both of these susceptibilities are dependent on the chosen EOSs. In the context of bulk viscosity, the parameter ${\cal C}$ is of importance due to its strong variation with density. On the other hand, the variation of ${\cal B}$ with density is minimal, which leads to the bulk viscosity coefficient $\zeta$ being largely unaffected by ${\cal B}$.

 iii. MURCA interaction rate:

In this study, we have focused on calculating the bulk viscosity of the baryonic medium in the presence of trapped neutrinos, particularly for the MURCA process. Initially, we derived semi-analytical expressions for the rates of two MURCA processes: $n+N\leftrightarrow N+p+e+\bar\nu_e$ and $N+p+e\leftrightarrow N+n+\nu_e$. These rates are functions of density, temperature, and $\mu_{\Delta}$. We neglected terms involving anti-neutrinos, as they are suppressed under the semi-degenerate condition $\mu_{i}> T$. To determine the chemical potential fluctuation, we solved an integro-differential equation, obtaining the general solution $\mu_{\Delta}/T$. The nature of $\mu_\Delta/T$ is found to be anharmonic and $\mu_{\Delta}/T<1$. By integrating $\mu_\Delta/T$ weighted with the cosine of the angular frequency over one oscillation period, we derive the bulk viscosity of the hadronic medium.

 iv. Lepton fraction dependence:

 In Fig.(\ref{vary_temp}), we present the temperature variation of $\zeta$. The plot demonstrates a resonant behavior, where the bulk viscosity exhibits resonance when the angular frequency of the merged object matches the interaction rate of the MURCA process.  Notably, we observe that for lower lepton fractions, $\zeta$ is more pronounced, while for higher lepton fractions, $\zeta$ is reduced. The reason for this trend is that higher lepton fractions lead to an increase in the feedback term in the integro-differential equation, resulting in smaller values of $\mu_{\Delta}/T$.  As $\mu_{\Delta}/T$ decreases  $\zeta$  decreases consequently.

 v. Temperature dependence:

 $\zeta$ in neutrino trapped baryonic matter  as a function of temperature  at a fixed oscillation frequency of $8.4$ kHz shows resonant behaviour. In neutrino transparent matter the maxima appears at lower temperature. The height of the maxima as well as its position  change in neutrino trapped matter.  The position of peak of the curve ($T_{\zeta_{max}}= \omega /(\tilde \Gamma {\cal B}))^{1/m}$) depends upon both interaction rate and nuclear susceptibility. The height of the peak depends upon the susceptibilities  $\zeta_{max}={\cal C}^2\tau/4\pi {\cal B} $. Hence, change in EoS and $\Gamma^{\leftrightarrow}$ change the height of the resonant curve and also the position of the peak towards higher temperature. Hence, $\zeta$ with trapped neutrino is more relevant in the context of binary neutron star merger.

 vi. Density dependence:
 
 The density dependence of $\zeta$ is primarily influenced by the density dependence of the susceptibilities. These susceptibilities are determined by the underlying EoS of the medium. Specifically, ${\cal B}$ shows a weak dependence on density, while ${\cal C}$ exhibits a strong dependence on density. As a result, the variation of bulk viscosity with density closely mirrors that of ${\cal C}$.

 vii. Time scale related to bulk viscosity:

 The characteristic time scale, which relies on bulk viscosity, isothermal compressibility, and angular frequency of oscillation, has been computed. For  $T=20$ MeV, $\tau$ lies within the range of $35-140$ milliseconds for NL3 (Fig.(\ref{time_scale})). Remarkably, this scale aligns with the survival time period of the compact object after merging. 

The present formulation of bulk viscosity in binary neutron star mergers establishes a connection between the dense matter found in binary neutron star mergers and the matter encountered in heavy-ion collisions. In both scenarios, dissipative processes such as viscosity play a vital role in defining the properties of dense matter generated at elevated temperatures and high densities. 
In the density regime considered here, hyperons might also appear \cite{Thapa:2021ifv} . Thus, the effect of hyperons on the EOS and the hyperon bulk viscosity might be significant \cite{Ofengeim:2019fjy}. However, the appearance of hyperons depends on the choice of optical potential depths of hyperons, which are still uncertain to some extent. In future study we   plan to extend our calculation to incorporate a diverse set of baryonic matter equations of state,
 as well as the quark-hadron mixed phase,   to obtain a more realistic representation of neutron star mergers.
\subsection*{Acknowledgments}
 S. Sarkar would  like to thank and acknowledge T. Mazumder for  fruitful discussions regarding various aspects of this work. 
\section{Appendix}
\label{App}
 \subsection{Beta Equilibration Rate}
 \label{int_rate}
In the Appendix we present the calculation of MURCA interaction rate in detail.  
The expression for equilibration rate can be written as, 
\bea
\Gamma^{\leftrightarrow}_{1,2}=\Gamma^{\rightarrow}_{1,2}-\Gamma^{\leftarrow}_{1,2}=AI_{1,2},
\eea 
 where, $A$ is the angular integral and $I$ is the energy integral. In the angular integration we consider the momentum of neutrons, protons and electrons within a  few $\sim k_BT$  of the Fermi energies. Neutrino momentum ($\sim k_BT/c$) is small in comparison to other momenta, hence neglected.  A more accurate treatment considering neutrino momentum in the angular delta function will be addressed in a future work.
 
 The angular momentum integration is evaluated as follows \cite{Shapiro, Yakovlev:2000jp},
 \bea
 A&=&\int_{j=1}^6d\Omega_j \delta^3(p_f-p_i)\nn\\
 &=&4\pi\int_{j=1}^5d\Omega_j \delta^3(p_f-p_j)\nn\\
 &=&\frac{2\pi(4\pi)^4}{p_{fn}p_{fN}p_{fN}'}\qquad \text{when}\quad p_{fn}>(p_{fe}+p_{fp}).
 \eea
In the above equation $p_{fn}$, $p_{fN}$, $p_{fN}'$, $p_{fp}$ and $p_{fe}$,  are the Fermi momenta of neutrons, initial spectator neutrons, final spectator neutrons, protons and electrons, respectively.  

The energy integral in Eq. \ref{eq:int_rate}, takes the following form:
 \bea
 I_1&=&\int p_n^2dp_np_N^2dp_Np_N'^2dp_N' p_p^2dp_pp_e^2dp_ep_{\bar \nu}^2dp_{\bar \nu}\nn\\
&&f_Nf_pf_ef_{\bar\nu}(1-f_N)(1-f_n)\delta(p'_N+p_p+p_e+p_{\bar\nu_e}-p_n-p_N)\nn\\
&&-f_nf_N(1-f_N)(1-f_p)(1-f_{\bar\nu_e})(1-f_e)\delta(p_n+p_N-p'_N-p_p-p_e-p_{\bar\nu_e}).
\label{en_int1}
\eea
 Our calculation of the MURCA phase space relies on a non-relativistic approximation for neutrons and protons, using $p_jdp_j = m_j^{\star}dE_j$ ($j = n, N, p$) to obtain the following expression \cite{Alford:2019qtm, Shapiro},
\bea
I_1&=&p_{fn}m_np_{fN}m_Np'_{fN}m_Np_{fp}m_pp_{fe}^2\int E_{\nu_e}^2 dE_{\nu_e}dE_{n}dE_{N}dE_{N}dE_{p}dE_{e}\nn\\
&&f_Nf_pf_ef_{\bar\nu_e}(1-f_N)(1-f_n)\delta(p'_N+p_p+p_e+p_{\bar\nu_e}-p_n-p_N)\nn\\
&&-f_nf_N(1-f_N)(1-f_p)(1-f_{\bar\nu_e})(1-f_e)\delta(p_n+p_N-p_p-p_e-p_{\bar\nu_e}-p'_N).
\eea
To perform the energy integral, we use the following substitutions in the two delta functions in the above equation as described below,
\bea
&& \delta(p_n+p_N-p'_N-p_p-p_e-p_{\bar\nu_e})\nn\\
&&= \delta(E_ n+E_ N-E'_ N-E_p-E_e-E_{\nu_e}-(\mu_n+\mu_n-\mu_n-\mu_p-\mu_e+\mu_{\nu_e})+(\mu_n-\mu_p-\mu_e+\mu_{\nu_e}))\nn\\
 &=&\delta(x_1+x_2+x_3+x_4+x_5+(-x_6 +\mu_{\Delta}/T))/T.
 \label{delta1}
\eea

In the above equation we substitute  $E_n$, $E_N$, $E'_ N$, $E_p$, $E_e$, $E_{\nu_e}$ with $x_i$ (i=n, N, p, e, $\nu_e$) using the relations  $x_1=\beta(E_n-\mu_n)$, $x_2=\beta(E_N-\mu_N)$, $x_3=-\beta(E'_N-\mu_N)$, $x_4=-\beta(E_e-\mu_e)$, $x_5=-\beta(E_p-\mu_p)$, $x_6=\beta(E_{\bar\nu_e}+\mu_{\nu_e})$ with $\Delta\mu=\mu_n-\mu_p-\mu_e+\mu_{\nu_e}$. In compact notation  we write the energy integral as,
\bea
&&I_1=
{\cal C}onst\hspace{.1cm}T^7\int dx_{\nu_e} x_{\nu_e}^2\int\Pi_{i=1}^5dx_i \nn\\
&&[(1+e^{x_i})^{-1}(1-f_{\bar\nu_e})-(1+e^{x_i})^{-1}f_{\bar\nu_e}]\delta(x_1+x_2+x_3+x_4+x_5+(-x_6 +\mu_{\Delta}/T))/T\nn\\
&&={\cal D}(I_{10}-I_2(\nu)-I_3(\nu)),
\label{I1}
\eea
${\cal C}onst=-p_{fn}m^{\star4}p_{fN}p_{fn}p_{fp}p_{fe}^2$. In the above equation the contribution arising from terms containing antineutrino distribution functions ($I_2(\nu)$ and $I_3(\nu)$) are suppressed as well as we have neglected the contributions coming from  the term $\mu_{\nu_e}/T $ which is less than one. Here, we have used $x_{\nu_e}=E_{\nu_e}/T$. 

For the other MURCA process $N+p+e\leftrightarrow N+n+\nu_e$, the the energy integral in the interaction rate $\Gamma_2^{\leftrightarrow}$  is performed in the following way,
\bea
&&I_2={\cal C}onst\hspace{.1cm}T^7\int dx_{\nu_e} x_{\nu_e}^2 \int\Pi_{i=1}^5dx_i \nn\\
&&\l[\l(1+e^{x_j}\r)^{-1}\l(1-f_{\nu_e}\r)-\l(1+e^{x_j}\r)^{-1}f_{\nu_e}\r]\delta \l( x_1+x_2+x_3+x_4+x_5+\l(-x_6 -\mu_{\Delta}/T\r)\r)/T\nn\\
&=&{\cal D}\l(I_{20}-I_4(\nu)-I_5(\nu)\r).
\label{I2}
\eea
 The second delta function has been written considering $x_1=\beta(E_n-\mu_n)$, 
 $x_2=\beta(E_p-\mu_p)$, $x_3=\beta(E_e-\mu_e)$, $x_4=-\beta(E_{n_e}-\mu_n)$, $x_5=-\beta(E'_N-\mu_N)$, $x_6=\beta(E_{\nu_e}-\mu_{\nu_e})$ \cite{Alford:2019kdw}.
In absence of neutrinos $I_2(\nu), I_3(\nu), I_4(\nu), I_5(\nu)$ vanish. 

Now, excluding the neutrino integral, we conduct all the remaining integrals in Eq.(\ref{I1}) and in Eq.(\ref{I2}) using the following technique \cite{1979ApJ...227..995E}, 
 \bea
&&\int \Pi_{i=1}^5dx_i \l(1+e^{x_i}\r)^{-1}\nn\\
 &&\delta\l( x_1+x_2+x_3+x_4+x_5+\l(-x_6 +\mu_{\Delta}/T\r)\r) +\delta\l( x_1+x_2+x_3+x_4+x_5+\l(-x_6 -\mu_{\Delta}/T\r)\r)\nn\\
&=&\Bigg[\frac{1}{1+e^{(-x_{6}+\mu_{\Delta}/T)}}\frac{1}{4!} ((-x_{6}+\mu_{\Delta}/T)^4+10\pi^2 (-x_{6}+\mu_{\Delta}/T)^2+9\pi^4)\nn\\
&+&\frac{1}{1+e^{(x_{6}+\mu_{\Delta}/T)}}\frac{1}{4!} ((x_{6}+\mu_{\Delta}/T)^4+10\pi^2 (x_{6}+\mu_{\Delta}/T)^2+9\pi^4)\Bigg].
\eea 
We employ the following method to evaluate the above integral, 
\bea 
\int \Pi_{i=1}^5dx_i \l(1+e^{x_i}\r)^{-1}\delta\l( x_i+y\r)=\frac{1}{\l(1+e^{-y}\r)}\frac{1}{4!}\l(y^4+10\pi^2 y^2+9\pi^4\r).
\label{neutrino_mom}
\eea 
Using Eqs.(\ref{en_int1}), (\ref{delta1}) and (\ref{I1}) the final expression for $\Gamma^{\leftrightarrow}_1$  becomes,
\bea
\Gamma_1^{\leftrightarrow}&\simeq &\tilde \Gamma T^7
\int dx_{\nu_e}x_{\nu_e}^2\frac{1}{1+e^{(-x_{6}+\mu_{\Delta}/T)}}\frac{1}{4!}\l[\l(\frac{\mu_{\Delta}}{T}-x_{6}\r)^4+10\pi^2\l(\frac{\mu_{\Delta}}{T}-x_{6}\r)^2+9\pi^4\r].
\eea
On the other hand, the final expression for $\Gamma_2^{\leftrightarrow}$  from Eqns.(\ref{en_int1}) and (\ref{I2}) becomes,
\bea
\Gamma_2^{\leftrightarrow}&=&\tilde \Gamma T^7
\int dx_{\nu_e}x_{\nu_e}^2
\l(\frac{1}{1+e^{(-x_{\nu_e}+\mu_{\nu_e}/T)}}\r)\l(\frac{1}{1+e^{(-x_{6}-\mu_{\Delta}/T)}}\r)\nn\\
&&\frac{1}{4!}\l[\l(\frac{\mu_{\Delta}}{T}+x_{6}\r)^4+10\pi^2\l(\frac{\mu_{\Delta}}{T}+x_{6}\r)^2+9\pi^4\r]\nn\\
&-&\l(\frac{1}{1+e^{(x_{\nu_e}-\mu_{\nu_e}/T)}}\r)\l(\frac{1}{1+e^{(-x_{6}-\mu_{\Delta}/T)}}\r)\frac{1}{4!}\l[\l(\frac{\mu_{\Delta}}{T}+x_{6}\r)^4+10\pi^2\l(\frac{\mu_{\Delta}}{T}+x_{6}\r)^2+9\pi^4\r],\nn\\
\eea
The final expression for MURCA equilibration rate from above two equations then  becomes,
\bea
\Gamma^{\leftrightarrow}&=&\Gamma_1^{\leftrightarrow}+\Gamma_2^{\leftrightarrow}\simeq\tilde \Gamma T^7
\int dx_{\nu_e}x_{\nu_e}^2\frac{1}{1+e^{(-x_{6}+\mu_{\Delta}/T)}}\frac{1}{4!}\l[\l(\frac{\mu_{\Delta}}{T}-x_{6}\r)^4+10\pi^2\l(\frac{\mu_{\Delta}}{T}-x_{6}\r)^2+9\pi^4\r]\nn\\
&+&\l(\frac{1}{1+e^{(-x_{\nu_e}+\mu_{\nu_e}/T)}}\r)\l(\frac{1}{1+e^{(-x_{6}-\mu_{\Delta}/T)}}\r)\frac{1}{4!}\l[\l(\frac{\mu_{\Delta}}{T}+x_{6}\r)^4+10\pi^2\l(\frac{\mu_{\Delta}}{T}+x_{6}\r)^2+9\pi^4\r]\nn\\
&-&\l(\frac{1}{1+e^{(x_{\nu_e}-\mu_{\nu_e}/T)}}\r)\l(\frac{1}{1+e^{(-x_{6}-\mu_{\Delta}/T)}}\r)\frac{1}{4!}\l[\l(\frac{\mu_{\Delta}}{T}+x_{6}\r)^4+10\pi^2\l(\frac{\mu_{\Delta}}{T}+x_{6}\r)^2+9\pi^4\r],
\eea
where, $\tilde{\Gamma}=-4.68\times 10^{-19.0}\left(\frac{x_p n_B}{n_0}\right)^{\frac{1}{3}} \left(\frac{m^{\star } } {m}\right)^4\,\,{\rm MeV}^{-3}$. 

\begin{figure}[h]
\includegraphics[width=0.6\textwidth,clip=]{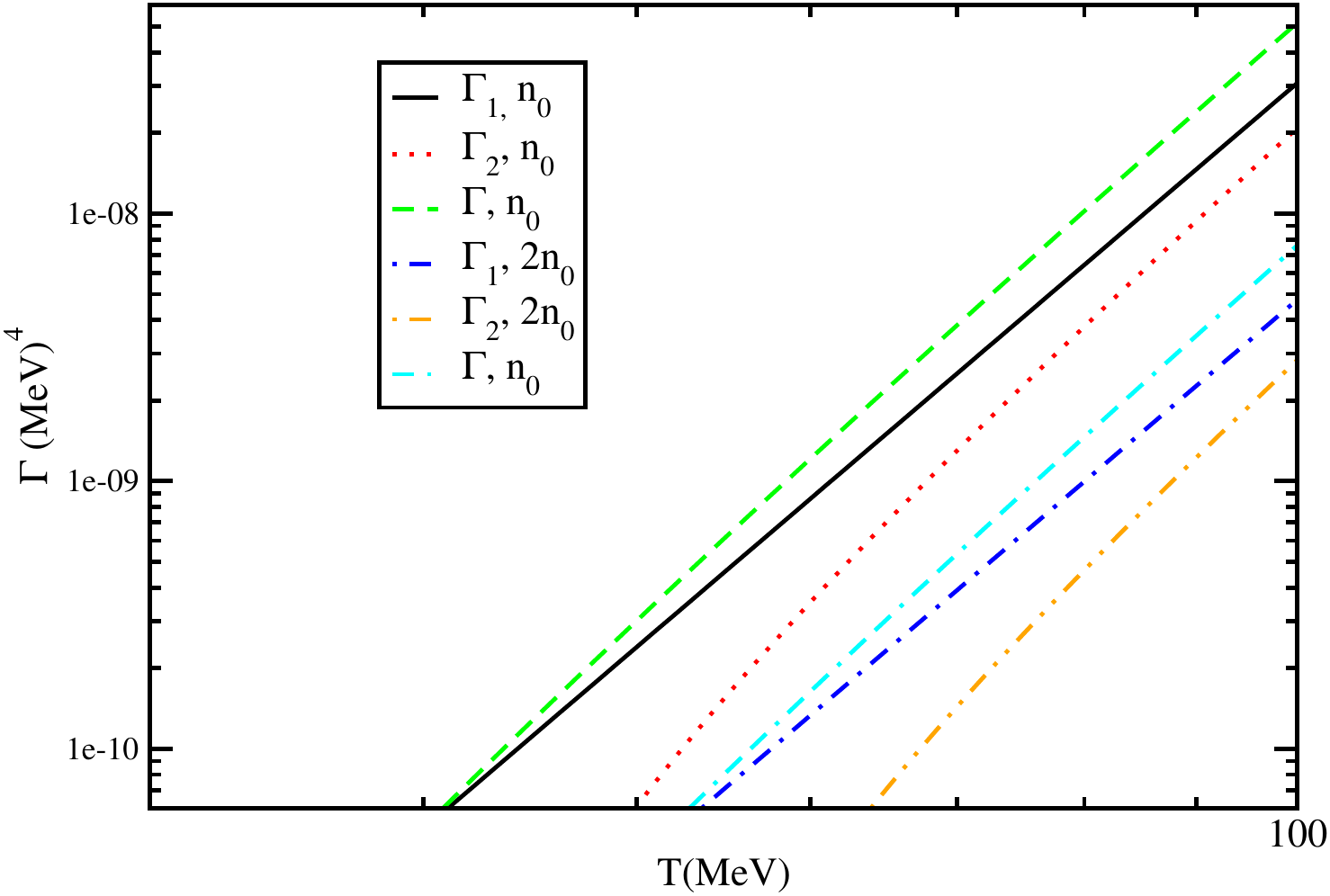}
   \caption{Variation of MURCA interaction rate with temperature for DDME2 equation of state for densities $n_B=n_0$, and $n_B=2n_0$, and considering $\mu_{\Delta}=0$.}
   \label{chem_pot_DDME2}
 \end{figure}
 We now present the variation in the MURCA interaction rate with temperature for $\Gamma_1^{\leftrightarrow}$, $\Gamma_2^{\leftrightarrow}$ and $\Gamma^{\leftrightarrow}$ at two different nuclear densities, $n_B = n_0$ and $n_B = 2n_0$. The plot is generated using the DDME2 EoS at a lepton fraction of $Y_l = 0.4$ (Fig.\ref{chem_pot_DDME2}). Black solid line corresponds to $\Gamma_1^{\leftrightarrow}$ for the density $n_B=n_0$, red dotted and green dashed lines correspond to $\Gamma_2^{\leftrightarrow}$ and total $\Gamma^{\leftrightarrow}$ respectively for the same density. Blue dot dased curve, orange double dot-dashed curve and cyan double dashed dotted curves correspond to $\Gamma_1^{\leftrightarrow}$, $\Gamma_2^{\leftrightarrow}$ and $\Gamma^{\leftrightarrow}$ respectively for the density $n_B=2n_0$. In the neutron decay MURCA process we have neglected the terms containing antineutrino distribution function. Both $\Gamma_1^{\leftrightarrow}$ and $\Gamma_2^{\leftrightarrow}$ shows power law variation with the temperature.

\subsection{Nuclear Equation of State}
\label{EOS}
We consider a medium of  nuclear matter consisting of n, p, e and $\nu_e$. 
We obtain chemical potentials of constituent baryons and leptons from both NL3  and DDME2 EoS.
The susceptibilities defined as ${\cal B}$ and ${\cal C}$  are  given below,
 \bea
{\cal B} &=&\frac{1}{\bar n_B}
\left(\frac{\partial \mu_{\nu_e}}{\partial x_n}\Big|_{n_B}+\frac{\partial \mu_{n}}{\partial x_{n}}\Big|_{n_B}-
\frac{\partial \mu_p}{\partial x_n}\Big|_{n_B}-
\frac{\partial  \mu_e}{\partial x_n}\Big|_{n_B}\right),\\
 {\cal C}&=&\bar n_B
\left(\frac{\partial \mu_{\nu_e}}{\partial n_B}\Big|_{x_n}
+\frac{\partial \mu_n}{\partial n_B}\Big|_{x_n}-
\frac{\partial \mu_p}{\partial n_B}\Big|_{x_n}-
\frac{\partial  \mu_e}{\partial n_B}\Big|_{x_n}\right).
\label{suscep}
\eea
Here ${\cal B}$ is the “beta-off-equilibrium–proton-fraction” susceptibility. It provides a  measure of  how the out-of-beta-equilibrium
chemical potential is related to  variation in the proton
fraction, whereas,  ${\cal C}$  is the “beta-off-equilibrium–baryon-density” susceptibility. ${\cal C}$ measures  the  variation of off-equilibrium
chemical potential to variation in the baryon
density at fixed proton fraction.

In the above two expressions $\mu_n$, $\mu_p$, $\mu_e$, $\mu_{\nu_e}$
 are the chemical potentials of neutrons, protons, electrons and electron neutrinos respectively. From relativistic mean field model the chemical potentials are given by,
\bea
\mu_n&=&\sqrt{m^{\star 2}+(3\pi^2x_n n
_B)^{2/3}}+g_{\omega}\omega_0-\frac{1}{2}g_{\rho}\rho_{03},\nn\\
\mu_p&=&\sqrt{m^{\star 2}+(3\pi^2x_p n_B)^{2/3}}+g_{\omega}\omega_0+\frac{1}{2}g_{\rho}\rho_{03}\nn\\
\mu_e&=&(3\pi^2x_p n_B)^{1/3},\nn\\
\mu_{\nu_e}&=&(3\pi^2\bar x_{\nu_e} n_B)^{1/3},
\label{chem_pot}
\eea
where $x_p$, $x_n$ and $\bar x_{\nu_e}$ are the proton, neutron and neutrino fractions respectively. They are related {\em via} the relations $x_p=1-x_n$, $\bar x_{\nu_e}=Y_l-(1-x_n)$, where, $Y_l$ is the lepton fraction. The number densities of electrons ($n_e$) and electron neutrinos ($n_{\nu_e}$) are linked to the lepton fraction through the relation $n_e+n_{\nu_e}=n_{L_t}=n_BY_l$. The effective mass of the nucleons is given by $m^{\star}=m-g_\sigma \sigma$, and $g_{\sigma}$, $g_\omega$ and $g_{\rho}$ are the couplings of the $\sigma$, $\omega$ and $\rho$ mesons with nucleons. 

 The final form of the beta disequilibration susceptibilities are given below,
 \bea
{\cal B}&=&B_0 +\frac{g_{\rho}^2}{m_{\rho}^2} - \frac{m^{\star 2}g_{\sigma}^2}{{\tilde m_{\sigma}^2\cal D}}\l(\frac{1}{E_{fn}} -\frac{1}{E_{fp}} \r)^2,\nn\\
{\cal C}&=&C_0 -\frac{g_{\sigma}^2m^{\star 2}}{{\tilde{m_{\sigma}}^2\cal D}}\l(\frac{1}{E_{fn}}-\frac{1}{E_{fp}}\r)\l(\frac{\mu_p}{E_{fp}}+\frac{\mu_n}{E_{fn}}\r)-\frac{1}{2}(\mu_p-\mu_n) .
\label{susceptibility}
\eea
In the above two equations $D$, $B_0$ and $C_0$ are given below,
\bea
D&=&1+\frac{g_{\sigma}^2}{\tilde m_{\sigma}^2}\l[\l(\frac{p_{fn}^3+3 m^{\star 2} p_{fn}}{E_{fn}}
-3m^{\star 2}\ln\Big|\frac{E_{fn}+p_{fn}}{m^{\star }}\Big|\r)+
\l(\frac{p_{fp}^3+3m^{\star 2}p_{fp}}{E_{fp}}-3m^{\star 2}\ln\Big|\frac
{E_{fp}+p_{fp}}{m^{\star }}\Big|\r)\r],\nn\\
C_0&=&\frac{1}{3}\l(\frac{p_{f n}^2}{E_{fn}}+\frac{p_{f\nu_e}^2}{E_{f\nu_e}}-\frac{p_{f p}^2}{E_{fp}}-\frac{p_{f e}^2}{E_{fe}}\r),\nn\\
B_0&=&\frac{1}{3}\l(\frac{1}{p_{f n}E_{fn}}+\frac{1}{p_{f \nu_e}E_{f\nu_e}}+\frac{1}{p_{f p}E_{fp}}+\frac{1}{p_{f e}E_{fe}}\r).
\label{suscep2}
\eea
Here, $\tilde m_{\sigma}^2=m_{\sigma}^2+2b_{\sigma}\sigma+3c_{\sigma}\sigma^2$ with $b_\sigma$ and $c_\sigma$ being the self-couplings of $\sigma$ meson \cite{Nandi:2020luz}, and $E_{fn}$, $E_{fp}$ and $E_{f\nu_e}$ are the  Fermi energy of neutrons, protons and electron neutrinos respectively. The expressions are similar to the equations given in Ref.\cite{Alford:2019kdw}. 

\bibliographystyle{abbrv}


\bibliographystyle{apsrev4-1}
\bibliography{main}

\end{document}